%% file: main.tex
\documentclass[runningheads]{llncs}
\usepackage[T1]{fontenc}
\usepackage{enumitem}

\usepackage{graphicx}
\usepackage{amsmath}
\usepackage{amssymb}
\usepackage{algorithm2e}
\usepackage{ifthen}
\usepackage{amsfonts,bbm}
\usepackage{url}
\usepackage{mathtools}
\usepackage{todonotes}
\usepackage{chngcntr}
\usepackage{nicefrac}
\usepackage{hyperref}
\usepackage{subdepth}
\usepackage{forest}
\usepackage{tikzducks}
\usepackage{etoolbox}
\usepackage{hhline}
\usetikzlibrary{shapes.geometric,positioning}
\usetikzlibrary{decorations.pathmorphing,shapes}
\tikzset{elli/.style={ellipse,draw}}

\newtheorem{defn}{Definition}
\newtheorem{rem}{Remark}
\newtheorem{prop}{Proposition}

\newcommand{\cupdot}{\mathbin{\mathaccent\cdot\cup}}
\usepackage{tikz}
\usepackage{subcaption}
\usetikzlibrary{arrows,automata,calc}

\newcommand{\reals}{{\mbox{\bf R}}}

\begin{document}
\title{On Sybil-proofness in Restaking Networks\thanks{Authors in alphabetical order.}}
%
%
\author{Tarun Chitra\inst{1} \and
Paolo Penna\inst{2} \and
Manvir Schneider\inst{3}}
\authorrunning{}
%
\institute{Gauntlet \and IOG \and Cardano Foundation}
%
\maketitle              
\begin{abstract}
Restaking protocols expand validator responsibilities beyond consensus, but their security depends on resistance to Sybil attacks. 
We introduce a formal framework for Sybil-proofness in restaking networks, distinguishing between two types of attacks, one in which other Sybil identities are kept out of an attack and one where multiple Sybil identities attack. 
We analyze marginal and multiplicative slashing mechanisms and characterize the conditions under which each deters Sybil strategies. 
We then prove an impossibility theorem: no slashing mechanism can simultaneously prevent both attack types. 
Finally, we study the impact of network structure through random graph models: while Erd\H{o}s--R\'enyi networks remain Sybil-proof, even minimal heterogeneity in a two-block stochastic block model makes Sybil attacks profitable. 
These results reveal fundamental limits of mechanism design for restaking and highlight the critical role of network topology.
\end{abstract}

\section{Introduction}

Restaking protocols have recently emerged as an important primitive in proof-of-stake blockchains. They allow staked tokens to be reused across multiple services, such as decentralized oracles, bridges, and data-availability layers. By opting in to secure additional services with their stake, validators can earn extra rewards, while services benefit from the economic security of the base layer. This mechanism has gained traction in practice, with systems such as EigenLayer~\cite{eigen-wp} on Ethereum already attracting significant adoption.

However, restaking also introduces new attack surfaces. A central concern is the possibility of Sybil attacks, where a participant splits their stake across multiple identities to manipulate service outcomes or reduce penalties. For example, if an attack requires a 100 ETH commitment but the attacker controls 200 ETH, committing the entire amount would expose all 200 ETH to slashing. By creating two identities, the attacker can risk only 100 ETH in the attack while keeping the other 100 ETH safe, effectively shielding half of their stake from loss (see Figure~\ref{fig:Sybil-example}). While the base proof-of-stake protocol is typically Sybil-resistant—since splitting does not change the probability of block selection—restaking changes the incentive structure. Different services may impose partial slashing rules, and rewards may depend on heterogeneous service outcomes. In such settings, splitting one’s stake across multiple identities can strictly improve an adversary’s payoff.

\begin{figure}[t]
\centering
\begin{minipage}{0.45\linewidth}
\begin{verbatim}
Without Sybil splitting:

 Attacker stake = 200 ETH
 Attack requires = 100 ETH

 → All 200 ETH committed
 → All 200 ETH slashed 
\end{verbatim}
\end{minipage}\hfill
\begin{minipage}{0.45\linewidth}
\begin{verbatim}
With Sybil splitting:

 Attacker stake = 200 ETH
 Split: 100 ETH (attack) + 
        100 ETH (safe)

 → Only 100 ETH committed
 → At most 100 ETH slashed 
\end{verbatim}
\end{minipage}
\caption{Illustration of how Sybil splitting can shield stake. 
By dividing into multiple identities, an attacker risks only 
a fraction of their total stake instead of the full amount.}
\label{fig:Sybil-example}
\end{figure}

\subsection{Our Contributions}
In this work, we analyze the conditions under which restaking networks are Sybil-proof. Our contributions are twofold. First, we study simple splitting attacks in the presence of partial slashing rules. We identify two canonical attack types and compare the effectiveness of marginal versus multiplicative slashing. We show that the marginal rule prevents one type of split attack but not the other, whereas the multiplicative rule blocks the latter but not necessarily the former. Second, we extend the analysis to random network models. We show that in homogeneous Erdős–Rényi graphs, splitting strictly decreases utility, while in heterogeneous stochastic block models splitting may increase the chance of adversarial success.

A central finding of this paper is an impossibility theorem: 
no slashing mechanism can simultaneously prevent both types of attacks.
This result identifies a fundamental trade-off in the design of restaking protocols: mechanisms that deter one class of attack inevitably leave the other exploitable. 
It complements our positive results for marginal and multiplicative schemes and motivates the study of network structure, 
where we show that heterogeneity itself can reintroduce profitable Sybil strategies.

Together, these results can help to design slashing rules and to understand how the structure of service networks influences Sybil-proofness. They highlight trade-offs between fairness, efficiency, and security in the design of restaking protocols.

\subsubsection{Contributions Summary and Roadmap.}
Our main contributions are summarized as follows:
\begin{itemize}[nosep]
    \item We formalize Sybil-proofness in restaking networks, distinguishing Type~I (only one Sybil identity attacks) and Type~II (multiple Sybils attack) attacks. 
    \item We analyze marginal and multiplicative slashing mechanisms and derive conditions under which each deters Sybil attacks. 
    \item We prove an impossibility theorem: no slashing mechanism can simultaneously prevent both attack types. 
    \item We study the role of network structure via random graph models. Erd\H{o}s-R\'enyi networks are Sybil-proof, whereas minimal heterogeneity in a two-block SBM reintroduces profitable Sybil strategies. 
\end{itemize}

\paragraph{Paper structure.} 
Section~\ref{sec:lit} discusses related literature. Section~\ref{sec:model} introduces the model. Section~\ref{sec:marginal} analyzes a marginal slashing mechanism, and Section~\ref{sec:multiplicative} a multiplicative mechanism. In Section~\ref{sec:impossibility} we discuss an impossibility result. Section~\ref{sec:sbm} studies Sybil-proofness in random graph models. Section~\ref{sec:conclusion} concludes.

\section{Related Literature}\label{sec:lit}
Restaking strengthens new services by borrowing economic security from a base chain, but it also couples failure modes across services. Durvasula and Roughgarden~\cite{durvasula2024robust} study such cascading risks and show that overcollateralization can make attacks unprofitable, even when failures propagate. Chitra and Pai~\cite{chitra2024much} analyze incentive designs that push operators to rebalance after shocks, limiting spillovers without heavy collateral requirements.

These works focus on systemic robustness under shared security. Our focus is complementary: strategic stake splitting (Sybil attacks). Splitting is largely neutral in base PoS selection, but partial slashing and multi-service heterogeneity can change incentives. Bar-Zur and Eyal~\cite{bar2025elastic} propose elastic restaking where stake can stretch after slashing. They show how controlled splitting can improve Byzantine robustness of the base layer, but do not analyze when splitting creates an advantage for attackers under e.g. partial slashing across services. Our optimization-based derivation connects to classical mechanism design with convex penalties and minimal intervention principles: we minimize total slashing subject to feasibility constraints, yielding KKT multipliers that act as service “prices.” This ties to Sybil-proof design ideas where penalties are linear in used stake and invariant to identity splits (cf. early work on Sybil-proofness and rational protocol design~\cite{sybil2007rational,clarke2005incentives}) and to collusion-resistant mechanism design~\cite{conitzer2010limits,abraham2006collusion}.

We contribute on this gap by (i) comparing marginal vs. multiplicative partial-slashing rules for Sybil-proofness, and (ii) showing how network structure matters: in homogeneous settings splitting is worse, while in heterogeneous settings it can be strictly beneficial.

\section{Model}\label{sec:model}
We follow the notation from \cite{durvasula2024robust,chitra2024much}. There is a set of services $S$ and a set of operators $V$. An edge $(s,v) \in E \subset S \times V$ if node operator $v$ is restaking for service $s$. $\sigma:V \to \reals_+$ is a function that maps each node $v\in V$ to the amount of stake $v$ has in the network. We use the notation $\sigma_v:=\sigma(v)$. $\pi:S \to \reals_+$ is a function that maps each service $s\in S$ to the maximum profit from attacking that can be realized for $s$. We use the notation $\pi_s:=\pi(s)$. $\alpha: S \to [0, 1]$ is a function that maps each service $s\in S$ to the threshold percentage of stake that needs to collude to attack the service $s$.
For a node operator $v$, we define its neighborhood (or boundary) as $\partial v = \{ s : (s,v) \in E \}$.
Similarly, for a service $s$, we define its neighborhood as $\partial s = \{ v : (s,v) \in E\}$.
For each service, we define the total stake that is restaked for service $s$, $\sigma_{\partial s}$ as
\[
   \sigma_{\partial s} = \sum_{v : (s,v) \in E} \sigma_v
\]
For any set $D \subset V$ or set $A \subset S$, we will slightly abuse notation and write
\begin{align*}
\sigma_D = \sum_{v \in D} \sigma_v && \pi_A = \sum_{s \in A}\pi_s
\end{align*}
$f$ is a function such that $f(\pi, A) = \pi_A$, later we might allow for different functions.

\begin{defn}[$f$-attack]
A restaking graph $G$ has an $f$-\emph{attack} at $(A,B) \subset S \times V$ if it is profitable and feasible:\footnote{These conditions were originally identified in the EigenLayer whitepaper~\cite{eigen-wp} for the special case that $f(\pi, A) = \pi_A$.}
\begin{align}
    f(\pi, A) &> \sum_{v\in B}\sigma_v = \sigma_B  & \text{\emph{(Profitability)}}\label{eq:costly-profit}\\
    \forall s \in A: \;\; \sum_{v \in B \cap \partial s} \sigma_v &\geq \alpha_s \sum_{v \in \partial s} \sigma_v = \alpha_s \sigma_{\partial s} & \text{\emph{(Feasibility)}} \label{eq:feasibility}
\end{align}
\end{defn}

\paragraph{Previous Analysis~\cite{chitra2024much}} Previous research has examined scenarios involving an $f$-attack $(A,B)$, where the total slashed amount is $\sigma_B$. In such cases, attackers lose their entire stake, even if only a portion of their stake would have been sufficient to meet the feasibility threshold for the attack.

\begin{defn}[Sybil]
    An operator node can divide itself into multiple smaller nodes, called \textit{Sybils}, and distribute its stake across these nodes.
\end{defn}

\begin{example}[Simple Example]
    Assume that that in a restaking graph there is a service $s_1$ and a node $v_1$ and $(\{s_1\},\{v_1\})$ is an $f$-attack. That is, $v_1$ attacking $s_1$ is profitable ($\pi_{1}>\sigma_{1}$) and feasible ($\sigma_{1} \geq \alpha_1 \sigma_{1}$). Unless $\alpha_1=1$, the stake of $v$ is strictly greater than is needed for feasibility. By maintaining the edges, ${v_1}$ could split into two nodes restaking with $s_1$ with stakes $\sigma_{1}^1=\alpha_1 \sigma_{1}$ and $\sigma_{1}^2=(1-\alpha_1)\sigma_{1}$, respectively. Now, the attacker would only attack with the one node with stake $\sigma_{1}^1$, and would only be slashed $\sigma_{1}^1$, while in the original attack, the total $\sigma_{1}$ would have been slashed. 
    The immediate advantage in utility by creating a Sybil for attacker $v_1$ is $\sigma_{1}^2$. See Figure~\ref{fig:sybil-comparison}.
    \input{intro-example}
\end{example}

\begin{defn}[Sybil Attack]
An attack is called a \textit{Sybil attack} if it involves an attacker who has created Sybils. We distinguish two canonical types.
\begin{enumerate}
    \item \textit{Type~I}: only \emph{one} Sybil identity of an attacker participates in the attack; the others remain passive.
    \item \textit{Type~II}: \emph{more than one} Sybil identity participates in the attack\footnote{In practice, a service may have to first approve new nodes (cf. EigenLayer's updated slashing model), making some Type~II attacks infeasible.}.
\end{enumerate}
This captures the economic distinction we analyze below; when convenient we refer to a $k$-split in examples, but the classification remains Type~I vs Type~II.
\end{defn}

\begin{defn}[Sybil-proofness]
An attack is \textit{type $K$ Sybil-proof}, $K \in \{I,II\}$ if no attacker improves utility by creating type $K$ Sybils (for any number of splits $k\ge 2$).
\end{defn}

\begin{defn}[Stable attack]
    An attack $(A,B)$ is \textit{stable} if every attacker in $B$ contributes positive stake to all services in $A$ that they are connected to, i.e., no attacker is redundant with respect to the feasibility constraints in \eqref{eq:feasibility}.
\end{defn}

\paragraph{Utility and profitability under partial slashing.}
In sections on partial slashing, we evaluate profitability via attacker utilities: the total profit from the attacked services $f(\pi,A)$ is distributed proportionally to attackers' used stake, and slashing is computed by the chosen mechanism. An attack is profitable if the sum of attackers' utilities is positive under the mechanism.

\section{Marginal slashing mechanism}\label{sec:marginal}
In this section, we describe a \emph{marginal} (partial) slashing mechanism.
Let $(A,B)$ be a stable attack, i.e. every attacker is contributing to the attack. Let $2^A$ denote the power set of $A$, excluding the empty set.
For any $S \in 2^A$, denote by $B_S$ all attacking nodes that attack and only attack the services in $S$. We can write $B_S = \{v\in B \ | \ A \cap \partial v = S \}$.
That is, $\{B_S\}_{S \in 2^A}$ is a disjoint partition of the attacker set $B$. In particular, $\bigcup_{S \in 2^A} B_S = B$ and for any $S,S' \in 2^A$ with $S \neq S'$, we have that $B_S \cap B_{S'}=\emptyset$. (Note that some $B_S$ might be empty.) 
Furthermore, for all $v\in B_S$, $v$ attacks all services $s \in S$ and $v$ does not attack any $s'\in A\setminus S$. 

Let $S \in 2^A$ and for all $s \in S$ define
\begin{equation}\label{eq:cbss}
        c_{B_S}^s := \alpha_s \sigma_{\partial s} - \sum_{\substack{S'\subset A \\ S' \neq S \\ s\in S'}} \sigma_{B_{S'}}
\end{equation}
and the maximum over services in $S$
\begin{equation}
    c_{B_S} := \max_{s\in S} c_{B_S}^s.
\end{equation}

\paragraph{Slashed Amount.} The slashed amount for the set $B_S$ is
\begin{align}\label{eq:group_slash}
    \psi_{B_S} = \big[\,\sigma_{B_S} - |B_S|(\sigma_{B_S}-c_{B_S})\,\big]_+
\end{align}
Each operator $v\in B_S$ is slashed
\begin{equation}\label{eq:individual_slash}
    \psi_v = \big[\,\sigma_v - (\sigma_{B_S}-c_{B_S})\,\big]_+
\end{equation}
and note that $\sum_{v \in B_S} \psi_{v} = \psi_{B_S}$.
Note that the brackets $[\cdot]_+$ ensure non-negativity and imply $0\le \psi_v \le \sigma_v$. In the interior regime where $c_{B_S}\in(0,\sigma_{B_S}]$, the brackets are inactive and the algebra below applies unchanged.

\begin{prop}\label{prop:simple} The following holds true:\footnote{Proof in Appendix~\ref{app:proofs-marginal}.}
\begin{enumerate}
    \item[(i)] The marginal slashing mechanism in \eqref{eq:group_slash}-\eqref{eq:individual_slash} is Type~I Sybil-proof.
    \item[(ii)] The marginal slashing mechanism in \eqref{eq:group_slash}-\eqref{eq:individual_slash} is not Type~II Sybil-proof.
\end{enumerate}
    
\end{prop}

\paragraph{Special case. } We look at the special case where all groups consist of at most one operator, that is $\forall S\in 2^A, |B_S|\leq 1$. For each non-empty $B_S$, the group $B_S$ (which consists of only one operator $v$) is slashed $\psi_{B_S} = \psi_v  = c_{B_S}= \max_{s \in S} \alpha_s \sigma_{\partial s} - \sum_{S'\in 2^A \setminus S: s\in S'}\sigma_{B_{S'}}$ using Equations~\eqref{eq:cbss}-\eqref{eq:individual_slash}.

\begin{rem}
    In Proposition~\ref{prop:simple} we find that the simple slashing mechanism is not Type~II Sybil-proof. For Type~II Sybil-proofness we study a different mechanism---multiplicative mechanism, in the next section.
\end{rem}

\begin{example}\label{ex:one-overlap}
Consider the restaking graph $G = (S,V)$ in Figure~\ref{fig:example} with $S = \{s_1,s_2\}$ and $V = \{v_0,v_1,v_2,v_3,v_4\}$.
Assume there is an attack $(A,B)$ with $A=\{s_1,s_2\}$ with $\pi_1,\pi_2$ (large enough) and $\alpha_1 = 2/3, \alpha_2 = 1/2$, and $B=\{v_1,v_2,v_3\}$ with $\sigma_1=1,\sigma_2=1.5, \sigma_3 = 1$. Furthermore, there are two non-attacking operators $v_0,v_4$ with stake $1$ each. 
The neighborhoods are $\partial s_1 = \{v_0,v_1,v_2\}$ and $\partial s_2 = \{v_2,v_3,v_4\}$ and we have that $B_{\{1\}} = \{v_1\}, B_{\{2\}} = \{v_3\}, B_{\{1,2\}} = \{v_2\}$.\footnote{We use shorthand notation and write $B_{\{i\}}$ instead of $B_{\{s_i\}}$}
Next, note that the minimum stake required to attack $s_1$ is $\alpha_1 \sigma_{\partial s_1} = 2/3 \cdot 3.5 = 2.33$ and for $s_2$ attackers need $\alpha_2 \sigma_{\partial s_2} = 1/2 \cdot 3.5 = 1.75$. Then,
\begin{itemize}
    \item For the attacking node $v_2$:
    \begin{itemize}
        \item $c_{B_{\{1,2\}}}^{s_1} = 2.33 - \sigma_{B_{\{1\}}} = 2.33 - \sigma_1 = 2.33-1 = 1.33$
        \item $c_{B_{\{1,2\}}}^{s_2} = 1.75 - \sigma_{B_{\{2\}}} = 1.75 - \sigma_3 = 1.75-1 = 0.75$
        \item $c_{B_{\{1,2\}}} = \max \{c_{B_{\{1,2\}}}^{s_1},c_{B_{\{1,2\}}}^{s_2}\} = 1.33$
        \item Hence, $\psi_{v_2} = c_{B_{\{1,2\}}}$.
    \end{itemize}
    \item For the attacking node $v_1$:\\ $c_{B_{\{1\}}}=c_{B_{\{1\}}}^{s_1} = 2.33 - \sigma_{B_{\{1,2\}}} = 2.33 - \sigma_2 = 2.33-1.5 = 0.83$ and hence $\psi_{v_1} = c_{B_{\{1\}}}$.
    \item For the attacking node $v_3$: \\ $c_{B_{\{2\}}}=c_{B_{\{2\}}}^{s_2} = 1.75 - \sigma_{B_{\{1,2\}}} = 1.75 - \sigma_2 = 1.75-1.5 = 0.25$ and hence $\psi_{v_3} = c_{B_{\{2\}}}$.
\end{itemize}
The slashing profile for attackers under the marginal slashing mechanism is $(\psi_{v_1},\psi_{v_2},\psi_{v_3)} = (0.83,1.33,0.25)$ and there is no incentive to create Type~I Sybils. \input{ex}
\end{example}
Further examples are provided in in Appendix~\ref{app:examples}.

\section{Multiplicative slashing mechanism}\label{sec:multiplicative}
In this section, we analyze a \emph{multiplicative} (partial) slashing \textit{design}. The rule charges each attacking player $v\in B$ the same fraction $\phi$ of its stake, i.e., a slash of $\phi\,\sigma_v$.
For a single service $s$, we set
\begin{align}
    \phi_s = \alpha_s \, \frac{\sigma_{\partial s}}{\sigma_{B}} \ .
\end{align}
Equivalently, the rule \emph{retains} a common fraction $1-\phi_s$ of each attacker’s stake. Note that this design charges the threshold amount (proportional to $\alpha_s\,\sigma_{\partial s}$), rather than only the \emph{excess over threshold}. It is identity-invariant and yields closed-form utilities. See Lemma~\ref{lem:lambda-phi} for how this choice relates to the minimal-slashing program (in the single-service case, the minimal program would slash by $\lambda^* = 1-\phi_s$).
Thus $\phi_s\in[0,1]$.
\begin{prop}\label{prop:multiplicative-slashing}
    The multiplicative slashing mechanism is Type~II Sybil-proof for any number of Sybils $k\ge 2$.\footnote{Proof in Appendix~\ref{app:proofs-marginal}.}
\end{prop}

We further assume that the value of the attacked service(s) is shared among the attackers in a ``fair'' way, namely, proportionally to the (used) stake, 
\begin{align}\label{eq:value-share}
    \pi_s(\sigma_v) = \pi_s \cdot \frac{\sigma_v}{\sigma_{B}}. 
\end{align}

\subsection{Principled derivation via minimal slashing}
We motivate partial slashing as the \emph{least intervention} needed to restore feasibility. The protocol aims to deter deviations without over-penalizing honest or minimally-contributing attackers. Formally, among all slashing vectors that reduce the \emph{post-slash used attacking stake} on every attacked service to the target level, we pick one that minimizes the total slashed stake. This ensures: (i) feasibility restoration, (ii) no over-slashing beyond what is necessary, (iii) identity invariance under splits (linearity in stake), and (iv) transparency via per-service multipliers.
Consider variables $\{\psi_v\}_{v\in B}$ and the program
\begin{align*}
    \min_{\{\psi_v\}\ge 0} \quad & \sum_{v\in B} \psi_v \\
    \text{s.t.}\quad & \sum_{v\in B\cap \partial s} (\sigma_v - \psi_v) \;\le\; \alpha_s \, \sigma_{\partial s} \qquad (\forall s\in A). 
\end{align*}
At any optimum with a successful pre-slash attack (i.e., when $\sum_{v\in B\cap \partial s} \sigma_v > \alpha_s\,\sigma_{\partial s}$), the inequalities bind as equalities.
Let $\lambda_s\ge 0$ be Lagrange multipliers for the (binding) constraints. The KKT conditions imply that any optimal solution satisfies
\begin{equation*}
    \psi_v = \big(\max_{s\in A\cap \partial v} \lambda_s\big)\,\sigma_v,\quad \text{with }\ \lambda_s>0\ \text{only on binding services}.
\end{equation*}
Thus, each attacker’s slashing factor equals the maximum multiplier of the services they attack, i.e., the \emph{max-scheme}. This rules out additive (sum) schemes as non-minimal (they double count across overlapping services). The program is convex, admits optimal solutions, and its KKT conditions certify optimality and uniqueness of the multipliers under mild regularity.

\begin{lemma}[Single-service multiplier and relation to multiplicative factor]\label{lem:lambda-phi}
Consider a single attacked service $A=\{s\}$ with aggregate attacking stake $\sigma_B$ and total restaked stake $\sigma_T=\sigma_{\partial s}$. If $\sigma_B>\alpha_s\,\sigma_T$, the minimal-slashing program above slashes each attacker by the same fraction
\[
\lambda^* \;=\; 1 - \frac{\alpha_s\,\sigma_T}{\sigma_B}\,.
\]
Equivalently, it retains a common fraction $1-\lambda^* = \tfrac{\alpha_s\,\sigma_T}{\sigma_B}$ of every attacker’s stake. If the multiplicative design is defined to slash a common fraction $\phi_s := \tfrac{\alpha_s\,\sigma_T}{\sigma_B}$, then $\lambda^* = 1-\phi_s$.
\end{lemma}

\begin{prop}[Max-of-factors is componentwise minimal among factorized rules]\label{prop:minimal-max}
Fix an attack $(A,B)$ and suppose a slashing rule is identity-invariant and factorized, i.e., there exist nonnegative per-service factors $\{\lambda_s\}_{s\in A}$ such that each attacker $v\in B$ is charged
\[
    \psi_v = \big(g(\{\lambda_s\}_{s\in A\cap \partial v})\big)\,\sigma_v,
\]
for some aggregation function $g$ that is monotone in each argument. Among such rules that restore feasibility (i.e., drive post-slash used stake to the thresholds), the choice
\[
    g(\{\lambda_s\}) = \max_{s\in A\cap \partial v} \lambda_s
\]
is componentwise minimal: for any other admissible factorized rule with aggregation $\tilde g\neq \max$, its slashing vector $\tilde\psi$ satisfies $\tilde\psi_v\ge \psi_v$ for all $v\in B$, with strict inequality on some instances.
\end{prop}

\subsection{Single service}
Consider an attack $(A,B)$ with single service $A=\{s\}$, and let us consider an arbitrary operator $v\in B$ and its Sybil strategy with stake $x\in [0,\sigma_v]$ in the attack and $\sigma_v - x$ not in the attack, i.e. Type~I Sybil. Let further denote by $\sigma_{B'}=\sigma_{B}-\sigma_v$ the remaining stake in the attack and by $\sigma_T= \sigma_{\partial s}$ the total stake restaked with $s$. Then, we can express the fraction $\phi$ of partial slashing as a function of $x$,  
\begin{align}\label{eq:multiplicative-single-factor}
     \phi(x) := \alpha_s \cdot \frac{\sigma_T}{x + \sigma_{B'}}. 
\end{align}
The mechanism slashes every $v' \in B'$ an amount $\hat \sigma_{v'} = \phi(x) \cdot \sigma_{v'}.$
Assuming the value $\pi_s$ is shared among the attackers proportionally to the (used) stake \eqref{eq:value-share}, we can write the utility of $v$ as a function of her Sybil stake $x$ as follows:
\begin{align*}
    u_v(x) &= \pi_s \cdot \frac{x}{x + \sigma_{B'}} - x \cdot \phi(x) \stackrel{\eqref{eq:multiplicative-single-factor}}{=}
    \pi_s \cdot \frac{x}{x + \sigma_{B'}} - x \cdot \alpha_s \cdot \frac{\sigma_T}{x + \sigma_{B'}} \\ &= (\pi_s - \alpha_s \cdot \sigma_T) \cdot \left(\frac{x}{x + \sigma_{B'}}\right). 
\end{align*}

\begin{prop}\label{prop:multiplicative-single-service}
    For a single service, the multiplicative partial slashing mechanism is Type~I Sybil-proof. That is, for any profitable attack (according to this partial slashing scheme), no attacker improves her utility by using Sybils, under the proportional value redistribution  in \eqref{eq:value-share}.
\end{prop}

The feasibility requires $x$ such that $\phi(x)\leq 1$, meaning that with such an $x$ the attack is feasible (we do not need more stake from the others than what they actually have). Assuming the attack is feasible (for the original definition), that is, $\phi(\sigma_v)\leq 1$, then profitability implies that the optimal $x$ is the feasible $x=\sigma_v$. \footnote{If $q(0)>0$ due to background randomness, the Jensen step in Proposition~\ref{prop:erdösreny-simple} is replaced by $q(x/k)\le q(x)\,\frac{1-q(0)}{k}+q(0)$, which suffices for the same qualitative conclusion for fixed $q(0)<1$ and $k\ge 2$.}

\subsection{Two Services}
We consider a setting of two services that are being attacked, that is, let $(A,B)$ be an attack with $A=\{s_1,s_2\}$ and $B = B_{\{s_1\}} \cupdot B_{{\{s_2\}}} \cupdot B_{{\{s_1,s_2\}}}$ the disjoint union of attackers. In particular, $B_S = \{v\in B \ | \ A \cap \partial v = S\}$ for $S \in 2^A$. A player $v \in B_{{\{s_1,s_2\}}}$ may follow a Type~I Sybil strategy. In particular, $v$ can split his stake into $x$ and $\sigma_v-x$, where the latter is not participating in the attack, $x$ stake contributes to attacking $s_1$ and $s_2$.
Then, 
\begin{align}\label{eq:phi-by-service}
    \phi_1(x) = \alpha_1 \frac{\sigma_{T_1}}{x+\sigma_{B_1'}}\ , && \phi_2(x) = \alpha_2 \frac{\sigma_{T_2}}{x+\sigma_{B_2'}},
\end{align}
where the total stake attacking $s_i$ is $x+\sigma_{B_i'}$ and $\sigma_{T_i} = \sigma_{\partial s_i}$ for $i \in \{1,2\}$.
We next consider how to compute the combined partial slashing for a player attacking both services. The minimal-slashing derivation above yields the \emph{max-scheme}:
\begin{align}
    \phi_{both}(x) = \max(\phi_1(x), \phi_2(x)). \tag{max-scheme} \label{eq:max-scheme}
\end{align}
The utility of player $v$ is given as sum of the profits of both attacked services net the slashed stake, 
\begin{align}\label{eq:utility-two-services}
    u_v(x) &= \pi_1 \frac{x}{x+\sigma_{B_1'}} + \pi_2 \frac{x}{x+\sigma_{B_2'}} - x\cdot \phi_{both}(x). 
\end{align}

We have the following result.
\begin{prop}\label{prop:max-scheme}
    For $\pi_1$ and $\pi_2$ large enough, the multiplicative partial slashing mechanism with \eqref{eq:max-scheme} is Type~I Sybil-proof. That is, an attacker will attack with her full stake and not create a Type~I Sybil.
\end{prop}

\begin{rem}
    Slashing the ``maximum'' between the two slashed stakes as in \eqref{eq:max-scheme} for the attackers in the intersection, means that the attackers in one service only will be charged either the same proportion (in one service) or a smaller proportion (for the other service). This is the unique minimal (componentwise) solution among factorized, identity-invariant slashing rules that restore feasibility.
\end{rem}

\begin{corollary}[n-service extension]\label{cor:n-services}
Let $A=\{s_1,\dots,s_n\}$ and suppose $v$ attacks every service in $A$ with stake choice $x\in[0,\sigma_v]$ engaged at all services. Under the max-scheme and proportional sharing, if $(\pi_i-\alpha_i\,\sigma_{T_i})\ge 0$ for all $i\in\{1,\dots,n\}$, then any best response is $x^*_v=\sigma_v$ (no Type~I Sybil). If $(\pi_i-\alpha_i\,\sigma_{T_i})<0$ for all $i$, then any best response is $x^*_v=0$.
\end{corollary}

We illustrate the multiplicative slashing scheme in Example~\ref{ex:multiplicative-example} in Appendix~\ref{app:examples}.


For completeness, we also examined an additive partial-slashing scheme as an alternative to the max-scheme. The qualitative takeaway is that, when profits are large enough, it is also Type I Sybil-proof (attackers optimally use their full stake). Full derivations are provided in Appendix~\ref{app:additive-scheme}. 

\paragraph{Sharing-rule sensitivity.}
The single-service Type~I result (full participation when $\pi_s>\alpha_s\sigma_T$) hinges on proportional sharing at the attacked service. Two illustrative alternatives:
\begin{itemize}
    \item \textit{Winner-take-all per service.} Rewards $\pi_s$ accrue entirely to the smallest coalition that clears the threshold. Under the max-scheme, this can create non-monotone best responses and reintroduce Type~I incentives even when $\pi_s>\alpha_s\sigma_T$.
    \item \textit{Pooled profits across services.} If rewards across attacked services are pooled before splitting proportionally, the two-service best-response derivative replaces $(\pi_i-\alpha_i\sigma_{T_i})$ by $(\sum_j\pi_j-\alpha_i\sigma_{T_i})$, expanding the region where full participation is optimal and shrinking withholding incentives.
\end{itemize}
Thus, proportional per-service sharing is sufficient (not necessary) for the stated Type~I results; pooling strengthens them, winner-take-all weakens them. More details in Appendix~\ref{sec:new-utility}. A full taxonomy is left to future work.

\section{Limits of Slashing Rules: An Impossibility Result}\label{sec:impossibility}

This section motivates and formalizes a limitation of partial-slashing designs: under mild axioms, no single rule can simultaneously eliminate incentives for both Type~I and Type~II Sybil strategies across all environments. We first explain why such a limitation is natural, then state the properties one may reasonably require of slashing rules, define them precisely, and prove the impossibility.

\subsection{Why a trade-off may be inevitable}
Type~II strategies (multiple identities all participating) exploit \emph{per-identity} artifacts. A sound defense is to make slashing \emph{identity-invariant}, i.e., depend linearly on stake regardless of how it is split. However, in heterogeneous multi-service settings, a Type~I strategy (withholding some identities) can shift exposure away from tighter constraints while preserving exposure to looser ones, potentially improving utility whenever profits are uneven. This suggests a fundamental tension: removing Type~II incentives via identity invariance can open profitable Type~I options in heterogeneous environments.

\subsection{Three desiderata for slashing rules}
We identify three properties that are both conceptually natural and practically desirable.
\begin{itemize}
  \item \emph{Identity invariance}: Splitting a stake into multiple identities that behave identically must not change total slashing for that stake.
  \item \emph{Feasibility restoration}: After slashing, the retained (used) attacking stake must restore (or preserve) feasibility constraints on every attacked service.
  \item \emph{Non-exploitative}: No attacker is slashed a negative amount nor more than its own stake.
\end{itemize}

\subsection{Mathematical definitions}
Fix an attack $(A,B)$ with per-attacker stakes $(\sigma_v)_{v\in B}$ and service constraints indexed by $s\in A$. A slashing rule is a mapping $\mathcal R$ producing nonnegative slashes $\psi=(\psi_v)_{v\in B}$.
\begin{defn}[Identity invariance]
For any attacker $v$ and any split of $\sigma_v$ into $k\ge 2$ parts $\{\sigma_v^{(i)}\}_{i=1}^k$ with $\sum_i \sigma_v^{(i)}=\sigma_v$, and any attack profile in which the $k$ identities behave identically (same edges and choices), the rule satisfies
\[
\sum_{i=1}^k \psi_{v^{(i)}}\big(\mathcal R\big) = \psi_v\big(\mathcal R\big).
\]
\end{defn}
\begin{defn}[Feasibility restoration]
Let the post-slash used stake be $x_v := \sigma_v - \psi_v$. For every attacked service $s\in A$, the retained attacking stake meets the threshold
\[
\sum_{v\in B\cap \partial s} x_v \;\le\; \alpha_s\,\sigma_{\partial s}.
\]
In instances where the pre-slash attack is successful (i.e., the pre-slash used stake exceeds the threshold), minimal slashing drives the post-slash used stake \emph{to} the threshold, so the inequality binds at the optimum.
\end{defn}
\begin{defn}[Non-exploitative]
For every attacker $v\in B$, the slash satisfies bounds
\[
0\;\le\;\psi_v\;\le\;\sigma_v.
\]
\end{defn}

\subsection{Impossibility theorem}
\begin{prop}[No universal rule eliminates both Sybil types]\label{prop:impossibility}
No slashing rule that satisfies identity invariance, feasibility restoration, and non-exploitative bounds can, for all attack instances and parameters, simultaneously eliminate incentives for both Type~I and Type~II Sybil strategies.\footnote{Proof in Appendix~\ref{app:proofs-impossibility}.}
\end{prop}

The takeaway is design-theoretic: rules that are split-proof against Type~II (identity-invariant) face inherent tension with Type~I in heterogeneous, multi-service environments. Mitigations include aligning reward-sharing and participation policies (e.g., pooled profits across services, service-level gating/approval, or dynamic penalties) to shrink the profitable withholding region.

\paragraph{Admission control (whitelisting) and caps.}
Suppose a service $s$ approves new nodes from a whitelist. Then a Type~II attacker cannot freely increase the number of participating identities on $s$. In our framework, this truncates the feasible split set and weakens the Type~II threat but does not remove the need for identity invariance in general multi-service attacks: an attacker can still split across other approved services. Proposition~\ref{prop:impossibility} continues to hold under whitelisting, as the proof constructs environments with pre-approved identities and relies on heterogeneous multipliers $(\lambda_s)$, not on unbounded $k$. Slashing caps (per-service maximum fractions) can reintroduce non-minimality and reduce Type~II deterrence by saturating at the cap; the max-scheme remains componentwise minimal \emph{subject to caps}, but strict Type~II proofness can fail if the binding cap is below the KKT multiplier.

\section{Sybil-proofness in the Stochastic Block Model (SBM)}\label{sec:sbm}

Sybil profitability is not only a property of the slashing rule, but also of the network topology. To capture this, we study Sybil-proofness in random graph models. Our approach uses a mean-field abstraction (full details in Appendix~\ref{app:sbm}) that allows us to replace intractable graph dependencies with probabilistic expectations.

We first show that in homogeneous Erdős–Rényi networks, splitting is strictly worse than using one identity (Proposition~\ref{prop:erdösreny-simple}). In contrast, introducing minimal heterogeneity via a two-block stochastic block model creates distinct “high-threshold” and “low-threshold” service environments. This asymmetry allows attackers to arbitrage across blocks, and Sybil splitting can strictly increase their probability of success (Proposition~\ref{prop:2sbm}).

The formal definitions of the SBM, clearance functions, and detailed derivations are deferred to the appendix; here we highlight the main results and their implications.

\subsection{Stochastic Block Model (SBM) for restaking networks}

Validators and services are partitioned into blocks, and the probability of an edge depends on the respective blocks. 
This allows us to interpolate between homogeneous Erd\H{o}s--R\'enyi graphs (one block) and heterogeneous networks with distinct service communities.  

To evaluate attack feasibility we define the \emph{clearance function}: for a service in block~$b$,
\begin{equation}
    q_b(y) \;=\; \Pr\!\left[\, y \;\text{clears the feasibility threshold of } s \in S_b \,\right],
\end{equation}
which gives the probability that an attacker with stake $y$ can successfully attack. 
Derivations using the mean-field approximation are given in Appendix~\ref{app:sbm}.  

An attacker committing stake $x$ then has success probability
\begin{equation}
    p(x) \;=\; \sum_b w_b \, q_b(x),
\end{equation}
while if the stake is split into $k$ Sybils of size $x/k$, the probability that at least one succeeds is
\begin{equation}
    p'(x;k) \;=\; 1 - \big(1 - p_k(x)\big)^k, 
    \qquad 
    p_k(x) = \sum_b w_b \, q_b(x/k).
\end{equation}

This framework lets us compare single-identity and Sybil strategies in both homogeneous and heterogeneous networks.  

\subsection{Sybil-proofness in the Erd\H{o}s-R\'{e}nyi random graph}
\paragraph{Nash best responses.} See Appendix~\ref{app:nash} for full statements and proofs of best responses and existence (single-service closed form; two-service monotonicity and existence conditions) under the max-scheme and proportional sharing.
A homogeneous network is one where all nodes are statistically equivalent and there are no coalition structures. In the context of our SBM framework, this corresponds to the simplest case where there is only one block for operators and one block for services ($R=1$).
Here every validator-service edge appears independently with probability $p$, and all services are statistically identical. 

In this setting, the clearance function $q(y)$ is strictly concave for sufficiently large networks (more details in Appendix~\ref{app:sbm}). 
Consequently, splitting stake strictly reduces the success probability:  
\begin{equation}
    p'(x;k) < p(x) \quad \text{for all } x>0,  k>1.
\end{equation}

\begin{prop}\label{prop:erdösreny-simple}
In Erd\H{o}s--R\'enyi networks, Sybil splitting is strictly dominated by a single-identity strategy.  
That is, Sybil attacks never improve success probability.
\end{prop}

\noindent
The intuition is that, since every service has identical difficulty, dividing stake only weakens each individual attempt, and the combinatorial advantage of multiple Sybils cannot compensate. 
This creates an illusion of security, however, the moment homogeneity is broken, this security guarantee can vanish. The following section will demonstrate this breakdown by introducing just one layer of heterogeneity via a two-block SBM.

\paragraph{Expected PNL under the max-scheme (ER).}
An expression for the expected PNL is deferred to Appendix~\ref{app:sbm}.

\subsection{Sybil-proofness in a Two-Block SBM}

The homogeneity of the Erd\H{o}s-R\'enyi model is the source of its Sybil-proofness. 
To find a vulnerability, we break this symmetry by constructing a two-block stochastic block model (SBM). 
This creates two distinct service environments: a ``high-threshold'' block with large expected total stake, and a ``low-threshold'' block with smaller expected total stake. 
An attacker connected to both faces a dilemma. 
A single identity must either commit a large stake (sufficient for the high-threshold block but excessive for the low-threshold block), or a small stake (well-suited for the low-threshold block but insufficient for the high-threshold block). 
By splitting, the attacker can present different stake profiles to each block, thereby arbitraging across them.

\begin{prop}
\label{prop:2sbm}
In a two-block SBM with heterogeneous thresholds, Sybil splitting can strictly increase the probability of success. 
Formally, there exist parameters for which $p'(x;k) > p(x)$ for some $k \geq 2$.
\end{prop}

This shows that minimal heterogeneity is sufficient to reintroduce a Sybil advantage. 
The formal proof and results on the expected PNL and on the optimal number of Sybils are given in Appendix~\ref{app:sbm}.

\paragraph{Discussion.}
The random graph analysis reinforces the impossibility result of Section~\ref{sec:impossibility}: 
no slashing mechanism can rule out Sybil attacks once heterogeneity is introduced. 
In homogeneous ER networks, concavity ensures Sybil-proofness, but even minimal structure in the interaction graph 
(as in the two-block SBM) suffices to reintroduce profitable Sybil strategies. 
This highlights that the limits identified in Section~\ref{sec:impossibility} are not only a feature of the mechanism, 
but also an inherent property of realistic network topologies.


\section{Conclusion}\label{sec:conclusion}

We studied Sybil-proofness in restaking protocols. 
Our analysis distinguished Type~I and Type~II attacks, 
and characterized the conditions under which marginal and multiplicative slashing schemes deter each type. 
A central result is an impossibility theorem showing that no mechanism can simultaneously prevent both attack types, establishing an inherent trade-off in restaking design. This motivates our analysis of random graph models: while homogeneous Erd\H{o}s--R\'enyi networks are Sybil-proof, even minimal heterogeneity in a two-block SBM suffices to reintroduce profitable Sybil strategies. 

Taken together, these results demonstrate both the power and the limitations of mechanism design for restaking. They highlight that ensuring Sybil resistance requires not only careful choice of slashing rules, but also attention to the structural properties of the validator-service interaction network. 

A natural direction for future work is to incorporate dynamic aspects such as repeated interactions, long-term rewards, and operator reputation. Another is to extend the analysis to more general network models and slashing schemes. Together, these directions can help clarify the security–efficiency trade-offs in designing robust restaking protocols.

\bibliographystyle{splncs04}
\bibliography{bib}

\appendix

\section{Proofs for the Marginal and Multiplicative Mechanisms}\label{app:proofs-marginal}

\paragraph{Proof of Proposition~\ref{prop:simple}.} Let $(A,B)$ be a stable attack. Let $\{B_S\}_{S\in 2^A}$ be a disjoint partition of $B$. Without Sybils, each group $B_S$ is slashed $\psi_{B_S}$ as in \eqref{eq:group_slash}, and each $v\in B_S$ is slashed $\psi_v$ as in \eqref{eq:individual_slash}. For (i): within a group $B_S$ an operator $v$ reducing its stake down to $\tilde\sigma_v\ge \psi_v$ keeps feasibility and the individual slash remains $\tilde\psi_v=\psi_v$, while its profit share weakly decreases, hence no incentive to Type~I. For (ii): splitting $\sigma_v$ into $\sigma_v^1,\sigma_v^2$ yields $\psi_v^1+\psi_v^2=\sigma_v-2(\sigma_{B_S}-c_{B_S})<\sigma_v-(\sigma_{B_S}-c_{B_S})=\psi_v$, thus Type~II reduces total slashing. 

\paragraph{Proof of Proposition~\ref{prop:multiplicative-slashing}.}
For Type~II: An operator $v$ with stake $\sigma_v$ is slashed $\phi\,\sigma_v$; splitting into $k$ identities with stakes summing to $\sigma_v$ leaves total slashing $\sum_i \phi\,\sigma_v^{(i)}=\phi\,\sigma_v$ unchanged. For Type~I in one service: the derivative \eqref{eq:multiplicative-single-factor}–\eqref{eq:value-share} implies $u_v'(x)=(\pi_s-\alpha_s\sigma_T)\,\frac{\sigma_{B'}}{(x+\sigma_{B'})^2}$, hence full participation if $\pi_s>\alpha_s\sigma_T$ and zero otherwise.

\paragraph{Proof sketch of Lemma~\ref{lem:lambda-phi}.}
By symmetry and linearity, any optimal solution must satisfy $\psi_v=\lambda\,\sigma_v$. The binding constraint $\sum_{v\in B}(\sigma_v-\psi_v)=\alpha_s\,\sigma_T$ pins $\lambda$ via $\lambda\,\sigma_B=\sigma_B-\alpha_s\,\sigma_T$, yielding the expression above.

\paragraph{Proof sketch of Proposition~\ref{prop:minimal-max}.}
Feasibility constraints are linear and separate by service. The Lagrangian dual uses one multiplier per service, and KKT complementary slackness implies $\psi_v=\big(\max_{s\in A\cap\partial v}\lambda_s\big)\sigma_v$ at any minimum. Any other monotone aggregation $\tilde g$ that maintains feasibility must satisfy $\tilde g\ge \max$ pointwise on instances where multiple service constraints bind; otherwise feasibility would be violated along some coordinate, establishing minimality and uniqueness within the factorized class.

\paragraph{Proof of Proposition \ref{prop:multiplicative-single-service}.}
Observe that 
\begin{align}
    \frac{\partial u_v}{\partial x} = (\pi_s - \alpha_s \cdot \sigma_T) \cdot \frac{\sigma_{B'}}{(x + \sigma_{B'})^2}, 
\end{align}
thus implying that the optimal strategy for $v$ is either to use no Sybil ($x=\sigma_v$ if $\pi_s > \alpha_s \cdot \sigma_T$), or to not take part of the attack ($x=0$ if $\pi_s < \alpha_s \cdot \sigma_T$). Note that the latter condition corresponds to the case the attack is \emph{not profitable} (even) under the partial slashing mechanism, 
\begin{align}
    \pi_s - \alpha_s \cdot \sigma_T = \pi_s - \phi(x) \cdot (x + \sigma_{B'})
\end{align}
as by definition $x + \sigma_{B'}$ is the overall stake in the attack, given the Sybil strategy $x$ of player $v$.  We have thus shown the following. 

\paragraph{Proof of Proposition \ref{prop:max-scheme}}
Let $v \in B_{{\{s_1,s_2\}}}$ be attacking both $s_1$ and $s_2$. As we are using the scheme in \eqref{eq:max-scheme}, the utility  \eqref{eq:utility-two-services} becomes
\begin{align}
    u_v(x) &= \pi_1 \frac{x}{x+\sigma_{B_1'}} + \pi_2 \frac{x}{x+\sigma_{B_2'}} - x\cdot \max(\phi_1(x),\phi_2(x)) \label{eq:max-slash}\\
    &= \left((\pi_1-\alpha_1 \sigma_{T_1})\frac{x}{x+\sigma_{B_1'}} + \pi_2 \frac{x}{x+\sigma_{B_2'}} \right) \mathbbm{1}_{\{\phi_1(x) \geq \phi_2(x)\}} +\\
    & \quad \left(\pi_1\frac{x}{x+\sigma_{B_1'}} + (\pi_2 - \alpha_2 \sigma_{T_2}) \frac{x}{x+\sigma_{B_2'}} \right) \mathbbm{1}_{\{\phi_1(x) < \phi_2(x)\}} \ . 
\end{align}
The condition in the indicator function $\phi_1(x)\geq \phi_2(x)$ is equivalent to
\begin{equation}\label{eq:two-services-phis-equal}
    x \geq \frac{\alpha_2 \sigma_{T_2} \sigma_{B_1'} - \alpha_1 \sigma_{T_1} \sigma_{B_2'}}{\alpha_1 \sigma_{T_1} -\alpha_2 \sigma_{T_2}}.
\end{equation}
The derivative of the utility function with respect to $x$ is
\begin{align}
    \frac{\partial u_v}{\partial x} &= \left((\pi_1-\alpha_1 \sigma_{T_1})\frac{\sigma_{B_1'}}{(x+\sigma_{B_1'})^2} + \pi_2 \frac{\sigma_{B_2'}}{(x+\sigma_{B_2'})^2} \right) \mathbbm{1}_{\{\phi_1(x) \geq \phi_2(x)\}} + \\
    & \quad \left(\pi_1\frac{\sigma_{B_1'}}{(x+\sigma_{B_1'})^2} + (\pi_2 - \alpha_2 \sigma_{T_2})\frac{\sigma_{B_2'}}{(x+\sigma_{B_2'})^2} \right) \mathbbm{1}_{\{\phi_1(x) < \phi_2(x)\}}.
\end{align}

Assume the attacker chooses $x \geq \frac{\alpha_2 \sigma_{T_2} \sigma_{B_1'} - \alpha_1 \sigma_{T_1} \sigma_{B_2'}}{\alpha_1 \sigma_{T_1} - \alpha_2 \sigma_{T_2}}$. Then, 
if $(\pi_1-\alpha_1 \sigma_{T_1}) > 0$, the derivative is positive, meaning that $x = \sigma_v$ maximizes the utility of the attacker $v$. That is, no Sybil and attacking with the full stake is the optimal strategy. If, however $(\pi_1-\alpha_1 \sigma_{T_1}) < 0$, there may be an incentive to do Sybils. In particular, the first order condition gives, 
\begin{equation}\label{eq:two-services-first-order}
    x = \frac{\sigma_{B_1'} - \sigma_{B_2'}\cdot A}{A-1}, \quad A:= \sqrt{\frac{\alpha_1\sigma_{T_1}-\pi_1}{\pi_2}\cdot \frac{\sigma_{B_1'}}{\sigma_{B_2'}}}.
\end{equation}

Similar for the case $x < \frac{\alpha_2 \sigma_{T_2} \sigma_{B_1'} - \alpha_1 \sigma_{T_1} \sigma_{B_2'}}{\alpha_1 \sigma_{T_1} - \alpha_2 \sigma_{T_2}}$. See Figure~\ref{fig:max-scheme-2services} for optimal choices when profits are large enough.

\begin{figure}[t]
    \centering
    \begin{tikzpicture}
    \draw[->,thick] (0,0) -- (7,0) node[below right] {$x$}; 

    \draw[thick] (0,-0.2) -- (0,0.2) node[above] {$0$}; 
    \draw[thick] (5,-0.2) -- (5,0.2) node[above] {$\sigma_v$}; 
    \draw[thick] (3,-0.2) -- (3,0.2) node[above] {$\frac{\alpha_2 \sigma_{T_2} \sigma_{B_1'} - \alpha_1 \sigma_{T_1} \sigma_{B_2'}}{\alpha_1 \sigma_{T_1} - \alpha_2 \sigma_{T_2}}$}; 

    \draw[->, thick] (0.3,-0.5) -- (2.7,-0.5) node[midway, below=2mm] {\small$\pi_2$ large enough}; 

    \draw[->, thick] (3.3,-1) -- (4.7,-1) node[midway, below=2mm] {\small $\pi_1$ large enough}; 
\end{tikzpicture}
    \caption{We have two regimes for smaller $x$ and for larger $x$. The optimal choice of $x$ given that it is small and $\pi_2$ large enough, is the rightmost point in the left regime. In the right regime, the rightmost point $\sigma_v$ is optimal for $\pi_1$ large enough.}
    \label{fig:max-scheme-2services}
\end{figure}

\paragraph{Proof sketch of Corollary~\ref{cor:n-services}.} The derivative of utility generalizes piecewise with $\phi_{\text{both}}(x)=\max_i \phi_i(x)$ and is a sum of nonnegative terms whenever every $(\pi_i-\alpha_i\sigma_{T_i})\ge 0$, implying monotonicity and the boundary optimum at $x=\sigma_v$. If all margins are negative, the derivative is nonpositive on each regime and $x=0$ maximizes utility.

\section{Examples}\label{app:examples}

\begin{example}\label{ex:one-overlap2}
    Recall Example~\ref{ex:one-overlap}. We replace $v_2$ by three operators $B_{\{1,2\}} = \{v_2^1,v_2^2,v_2^3\} $ with stakes 0.5, 0.75 and 0.25, respectively. Then, the slashing profile for these three nodes is $(\psi_{v_2^1},\psi_{v_2^2},\psi_{v_2^3)} = (0.33,0.58,0.08)$ while the slashing amounts for the other attackers stay as in the previous example. Note that the total slashed amount for the attackers attacking both services is 0.99, compared to the 1.33 when there is a single attacker attacking both services. If the three attackers are Sybils originating from $v_2$, then, $v_2$ saved 0.34 in slashing by creating Type~II Sybils. This example illustrates how the marginal slashing mechanism prone to Type~II Sybils.
\end{example}

\begin{example}\label{ex:multiplicative-example}
Recall Example~\ref{ex:one-overlap} with Figure~\ref{fig:example}.
Assume there is an attack $(A,B)$ with $A=\{s_1,s_2\}$ with $\pi_1,\pi_2$ and $\alpha_1 = 2/3, \alpha_2 = 1/2$, and $B=\{v_1,v_2,v_3\}$ with $\sigma_1=1,\sigma_2=1.5, \sigma_3 = 1$. Furthermore, there are two non-attacking operators $v_0,v_4$ with stake $1$ each.

The neighborhoods are as follows:
\begin{itemize}
    \item $\partial s_1 = \{v_0,v_1,v_2\}$
    \item $\partial s_2 = \{v_2,v_3,v_4\}$
    \item $v_2$ is attacking both, $v_1$ only $s_1$ and $v_3$ only $s_2$
\end{itemize}
Next, note that the minimum required stake to attack $s_1$ is $\alpha_1 \sigma_{\partial s_1} = 2/3 \cdot 3.5 = 2.33$ and for $s_2$ attackers need $\alpha_2 \sigma_{\partial s_2} = 1/2 \cdot 3.5 = 1.75$.

\begin{itemize}\label{ex:mult-syb-slash}
    \item No Sybil:
    With the multiplicative scheme (with full stakes), we have that 
    $\phi_1 = \frac{2.33}{1+1.5} = 0.932$ and $\phi_2 = \frac{1.75}{1.5+1} = 0.7$. Hence, the maximum of both is $\phi_1$, and the attacker attacking both services $v_2$ is slashed $\phi_1 \cdot 1.5 = 1.398$. $v_1$ is slashed the remainder to make the attack feasible: $2.33-1.398 = 0.932$. Same for $v_3$, it is slashed $1.75-1.398 = 0.352$. So the slashing profile for $(v_1,v_2,v_3)$ is $(0.932,1.398,0.352)$. 
    \item Type~I Sybil:
    However now, if $v_2$ would create a Sybil (1.4,0.1) and attack only with 1.4, then $\phi_1 = \frac{2.33}{1+1.4} = 0.97$, and therefore $v_2$ is slashed $\phi_1 \cdot 1.4 = 1.36$ and $v_1$ is slashed $0.97$, and $v_3$ is slashed $0.39$. Hence, by creating a Sybil, $v_2$ was slashed less and the other attackers more. So the slashing profile for $(v_1,v_2,v_3)$ is $(0.97,1.36,0.39)$. 
\end{itemize}
Total slashing without Sybil is $0.932+1.398+0.352=2.682$; with the Type~I split it is $0.97+1.36+0.39=2.72$. This aligns with the max-scheme’s componentwise minimality (Proposition~\ref{prop:minimal-max}): moving away from the max aggregation increases some components and can raise the total.
\end{example}

\section{Alternative additive scheme (sketch)}\label{app:additive-scheme}
For completeness, one can consider an additive scheme; we only sketch the main expressions and focus on the principled max-scheme.
\begin{align}
    u_v(x) &=  \left((\pi_1-\alpha_1 \sigma_{T_1})\frac{x}{x+\sigma_{B_1'}} + (\pi_2 -\alpha_2\sigma_{T_2}) \frac{x}{x+\sigma_{B_2'}} \right) \mathbbm{1}_{\{\phi_1(x) + \phi_2(x)\leq 1\}} +\\
    & \quad \left(\pi_1\frac{x}{x+\sigma_{B_1'}} + \pi_2  \frac{x}{x+\sigma_{B_2'}} - x\right) \mathbbm{1}_{\{\phi_1(x) + \phi_2(x)> 1\}} \ . 
\end{align}
The derivative is
\begin{align}
    \frac{\partial u_v}{\partial x} &= \left((\pi_1-\alpha_1 \sigma_{T_1})\frac{\sigma_{B_1'}}{(x+\sigma_{B_1'})^2} + (\pi_2 - \alpha_2 \sigma_{T_2}) \frac{\sigma_{B_2'}}{(x+\sigma_{B_2'})^2} \right) \mathbbm{1}_{\{\phi_1(x) + \phi_2(x) \leq  1\}} + \\
    & \quad \left(\pi_1\frac{\sigma_{B_1'}}{(x+\sigma_{B_1'})^2} + \pi_2\frac{\sigma_{B_2'}}{(x+\sigma_{B_2'})^2} -1\right) \mathbbm{1}_{\{\phi_1(x) + \phi_2(x)> 1\}}.
\end{align}
Assume that the attacker chooses $x$ such that $\phi_1(x) + \phi_2(x) \leq  1$. Then, if, for example, $(\pi_1-\alpha_1 \sigma_{T_1}) > 0$ and $(\pi_2-\alpha_2 \sigma_{T_2}) > 0$, the derivative is positive, meaning that $x = \sigma_v$ maximizes the utility of the attacker $v$. That is, no is Sybil and attacking with the full stake is the optimal strategy. 
In the other case, we need $\pi_1\frac{\sigma_{B_1'}}{(x+\sigma_{B_1'})^2} + \pi_2\frac{\sigma_{B_2'}}{(x+\sigma_{B_2'})^2} >1$ to ensure that no Sybils. This is again true if $\pi_1$ and $\pi_2$ are large enough.
We have shown the following.
\begin{prop}\label{prop:sum-scheme}
    For $\pi_1$ and $\pi_2$ large enough, the \emph{additive} partial-slashing scheme described above is Type~I Sybil-proof: an attacker’s best response is to use full stake (no Type~I Sybil).
\end{prop}

\section{Alternative utility function}\label{sec:new-utility}
    Consider Example~\ref{ex:mult-syb-slash} and consider the utility functions instead of only slashed stake. We observe that attacking with more stake is beneficial for the attacker $v_2$ as she receives a larger share of the profit.
    For example, let $\pi_1 = \pi_2 = 2$. By the definition of profitability \eqref{eq:costly-profit} the sum of profits ($\pi_1+\pi_2$) is equal to 4, while the sum of the slashed amounts will be below 3.5 (=$\sigma_{1}+\sigma_{2}+\sigma_{3}$).
\begin{itemize}
    \item Without Sybil ($x=\sigma_2$):
    \begin{align}
        u_2(x) &= \pi_1 \frac{x}{x+\sigma_{B_1'}} + \pi_2 \frac{x}{x+\sigma_{B_2'}} - x\cdot \max(\phi_1(x),\phi_2(x))\\
        &= 2 \frac{1.5}{2.5} + 2 \frac{1.5}{2.5} - 1.5 \cdot 0.932 \\ &= 1.002
    \end{align}
    \item With Type~I Sybil ($x<\sigma_2$):
    \begin{align}
        \Tilde{u}_2(x) &= \pi_1 \frac{x}{x+\sigma_{B_1'}} + \pi_2 \frac{x}{x+\sigma_{B_2'}} - x\cdot \max(\phi_1(x),\phi_2(x)) \\
        &= 2 \frac{x}{x+1} + 2 \frac{x}{x+1} - x\cdot \frac{2.33}{x+1}
\end{align}
and $\Tilde{u}_2(x)>u_2(x) \iff x>1.5$, which means that from a utility perspective there is no incentive for a Type~I Sybil.
\item More generally, for any $\pi_2$ and $\pi_1 = 2.33 - \pi_2$, any $x<1.5$ will lead to $\Tilde{u}_2(x)>u_2(x)$, however, $\pi_1+\pi_2=2.33$ way less than the total slashed amount.
\end{itemize}
Note that for node $v_1$ it may not be even worth attacking with this type of utility function. In particular for an attack with full stakes, 
\begin{align}\label{eq:neg-utility-problem}
    u_1(\sigma_1) &= \pi_1 \frac{\sigma_1}{\sigma_1+\sigma_2} - \sigma_1 \phi_1(\sigma_1) = 2 \frac{1}{2.5} - 0.932 = -0.132
\end{align}
In fact, $u_1(x)$ is decreasing in $x$ and would be maximized at $x=0$, i.e. not attacking.

How to go about the issue with negative utilities in Equation~\eqref{eq:neg-utility-problem}? Let $(A,B)$ be an attack. The utility of a player $v\in B$ is
\begin{align}
    u_v (x) = f(\pi,A) \frac{x}{x+\sigma_{B'}} - x \phi_v(x)
\end{align}
where $B' = B\setminus \{v\}$.

In the case of two services, $\phi_v (x) = \phi_{both}(x)$. The utility of node $v_1$ from Example~\ref{ex:mult-syb-slash} is positive. Indeed, 
\begin{equation}
    u_1(\sigma) = 4 \frac{1}{3.5} - 1 \cdot 0.932 \approx 0.21.
\end{equation}

\section{Proof for Impossibility Proposition}\label{app:proofs-impossibility}
Under identity invariance, total slashing for a given aggregate stake is independent of the number of identities. Consider two services with multipliers $\lambda_1>\lambda_2\ge 0$ and proportional sharing. For an attacker $v$ engaging $x$ at both services, $u_v(x)=\pi_1\frac{x}{x+\sigma'_{B_1}}+\pi_2\frac{x}{x+\sigma'_{B_2}}-\lambda_1 x$. For $\lambda_1$ large enough and $\pi_2$ sufficiently large (and with slack at $s_1$ maintained by others), $u_v'(x)<0$ at the current $x$, so withholding a small amount (Type~I) strictly improves utility while feasibility and bounds remain satisfied. Thus no rule satisfying the three properties can eliminate both types in all environments.

\paragraph{Proof of Proposition~\ref{prop:impossibility}.}
\emph{Step 1 (Type~II pressure forces identity invariance).} Any rule with per-identity thresholds or nonlinear aggregation can be gamed by splitting into sufficiently many identical identities that all participate, strictly reducing total slashing for fixed aggregate stake. Hence, to preclude Type~II gains \emph{universally}, identity invariance is necessary: total slashing must depend only on aggregate stake and not on the number of identities when behavior is identical.

\emph{Step 2 (With identity invariance, heterogeneous services induce Type~I incentives).} Consider two attacked services $s_1,s_2$ with proportional sharing of profits and feasibility-binding multipliers $\lambda_1>\lambda_2\ge 0$ (as in the KKT characterization of minimal slashing). Identity invariance implies that an attacker $v$ connected to both services is charged a factor equal to $\max(\lambda_1,\lambda_2)=\lambda_1$ on the stake it commits to the intersection. Let $\pi_1,\pi_2>0$ denote service profits allocated proportionally to used stake. If $v$ splits into two identities and withholds one identity on $s_1$ (Type~I), it reduces its slashed amount by $\lambda_1$ times the withheld stake, while losing only proportional profit on $s_1$ and preserving its exposure and profit share on $s_2$. Choosing parameters with $\pi_2$ sufficiently large and the background slack at $s_1$ sufficiently small (so that $\lambda_1$ remains binding via other attackers), the marginal reduction in slashing strictly dominates the marginal loss in profit, hence Type~I yields higher utility. Formally, write $u_v(x)=\pi_1\tfrac{x}{x+\Sigma'_1}+\pi_2\tfrac{x}{x+\Sigma'_2}-\lambda_1 x$ for $x$ stake engaged at both services (others fixed). For $\lambda_1>\pi_1\tfrac{\Sigma'_1}{(x+\Sigma'_1)^2}$ and $\pi_2\tfrac{\Sigma'_2}{(x+\Sigma'_2)^2}$ large enough, $\tfrac{\partial u_v}{\partial x}<0$ at the original $x$, showing a profitable reduction in $x$ (withholding) exists.

\emph{Step 3 (Non-exploitative bounds and feasibility are preserved).} The withholding move reduces $x$ but, by construction, keeps feasibility satisfied thanks to other attackers’ slack (feasibility restoration holds), and slashing remains in $[0,\sigma_v]$ (non-exploitative). Thus, any identity-invariant rule admits environments where Type~I is profitable.

Combining Steps 1–3, no slashing rule can satisfy the three properties and also eliminate both Sybil types uniformly across all environments.

\section{Nash Best Responses under the Max-Scheme}\label{app:nash}

In this appendix we characterize best responses and existence of equilibria under the multiplicative max-scheme with proportional sharing.

\subsection{Single service: best response and existence}
Consider a single service $s$ with total restaked stake $\sigma_T=\sigma_{\partial s}$ and threshold $\alpha_s\in(0,1]$. Let $B$ denote the set of attackers and $\sigma_B=\sum_{v\in B}\sigma_v$. Under the max-scheme, the slashing factor is
\[
\phi(x) = \alpha_s\,\frac{\sigma_T}{x+\sigma_{B'}},\quad \sigma_{B'}:=\sigma_B-\sigma_v,
\]
for a player $v$ choosing Sybil stake $x\in[0,\sigma_v]$. With proportional sharing, the utility of $v$ is
\[
u_v(x)=\pi_s\,\frac{x}{x+\sigma_{B'}}-x\,\phi(x) = (\pi_s-\alpha_s\sigma_T)\,\frac{x}{x+\sigma_{B'}},
\]
which is strictly increasing in $x$ if $\pi_s>\alpha_s\sigma_T$, strictly decreasing if $\pi_s<\alpha_s\sigma_T$, and flat at the boundary in the knife-edge case. Hence:
\begin{prop}
If $\pi_s>\alpha_s\sigma_T$, any best response is $x^*_v=\sigma_v$ (full participation). If $\pi_s<\alpha_s\sigma_T$, any best response is $x^*_v=0$ (no participation). If equality holds, any $x\in[0,\sigma_v]$ is optimal.
\end{prop}
Existence of a pure-strategy Nash equilibrium follows immediately: at a profitable service, the unique best response profile is full participation for each attacker; otherwise all attackers optimally choose zero.

\subsection{Two services: monotonicity and existence}
Let $A=\{s_1,s_2\}$ and suppose $v$ attacks both services with stake choice $x\in[0,\sigma_v]$ (engaged at both). Denote $\sigma_{B_i'}$ the aggregate stake of the other attackers at service $s_i$ and $\sigma_{T_i}=\sigma_{\partial s_i}$. The service-wise factors are
\[
\phi_i(x)=\alpha_i\,\frac{\sigma_{T_i}}{x+\sigma_{B_i'}},\quad i\in\{1,2\},
\]
and the max-scheme charges $\phi_{both}(x)=\max\{\phi_1(x),\phi_2(x)\}$. With proportional sharing, utility is
\[
u_v(x)=\pi_1\,\frac{x}{x+\sigma_{B_1'}}+\pi_2\,\frac{x}{x+\sigma_{B_2'}}-x\,\phi_{both}(x).
\]
The derivative is piecewise given by
\[
\frac{\partial u_v}{\partial x}=\begin{cases}
(\pi_1-\alpha_1\sigma_{T_1})\,\frac{\sigma_{B_1'}}{(x+\sigma_{B_1'})^2}+\pi_2\,\frac{\sigma_{B_2'}}{(x+\sigma_{B_2'})^2},& \phi_1(x)\ge \phi_2(x),\\[4pt]
\pi_1\,\frac{\sigma_{B_1'}}{(x+\sigma_{B_1'})^2}+(\pi_2-\alpha_2\sigma_{T_2})\,\frac{\sigma_{B_2'}}{(x+\sigma_{B_2'})^2},& \phi_1(x)<\phi_2(x).
\end{cases}
\]
Therefore the best response is monotone in the per-service net margins $(\pi_i-\alpha_i\sigma_{T_i})$: if both are positive, the derivative is positive on each regime and $x^*_v=\sigma_v$; if one is negative, the first-order condition yields an interior solution given by \eqref{eq:two-services-first-order} (when feasible); if both are negative, $x^*_v=0$.
\begin{prop}
Suppose $(\pi_1-\alpha_1\sigma_{T_1})\ge 0$ and $(\pi_2-\alpha_2\sigma_{T_2})\ge 0$. Then any best response is $x^*_v=\sigma_v$. If $(\pi_1-\alpha_1\sigma_{T_1})<0$ and $(\pi_2-\alpha_2\sigma_{T_2})<0$, any best response is $x^*_v=0$. In the mixed-sign case, if \eqref{eq:two-services-first-order} lies in $[0,\sigma_v]$ and respects the regime condition \eqref{eq:two-services-phis-equal}, it gives the unique interior maximizer; otherwise the maximizer is at the feasible boundary.
\end{prop}
Existence of a pure-strategy equilibrium follows from continuity of payoffs, compactness of strategy sets, and that each player’s best-response correspondence is nonempty and upper hemicontinuous (piecewise-smooth with boundary cases), so Kakutani applies.

\section{Sybil-proofness in the Stochastic Block Model (SBM)}\label{app:sbm}

This section finds that Sybil profitability is not an intrinsic feature of a given slashing mechanism alone, but is rather an emergent property that arises from the \textit{topology} of the validator-service interaction graph. The security of a restaking protocol is linked to the structure of the network it governs. When the network exhibits community structures, with dense clusters of interaction between certain groups of validators and services, arbitrage opportunities emerge which can be exploited by attackers.

This section studies the stochastic block model (SBM), a more expressive random graph model capable of representing community structures. Within the SBM framework, validators and services are partitioned into distinct blocks, and the probability of a connection (a restaking relationship) depends on the blocks to which the validator and service belong. This allows for the modeling of scenarios where, for instance, a "clique" of high-value services attracts a dense network of validators, while other services exist in sparser, less-connected regions of the graph. It is precisely this structural heterogeneity that a sophisticated Sybil attacker can leverage to their advantage, turning a seemingly secure protocol into a profitable target.

\paragraph{PNL Maximization.}
An attacker is modeled as a rational agent seeking to maximize their expected utility. The profit-and-loss (PNL) for an attacker $v$ with attacking stake $x_v$ targeting a set of services $A$ is given by the total rewards gained minus the total stake slashed:
\begin{equation}
PNL_v(x_v, x_{-v}, A) = \sum_{s \in A \cap \partial v} \pi_s \frac{x_v}{\sigma_{B_s}(x_v, x_{-v})} - x_v \cdot \max_{s' \in \partial v \cap A} \phi_{s'}(x_v, x_{-v})
\end{equation}
Here, $\pi_s$ is the profit from successfully attacking service $s$, $\sigma_{B_s}$ is the total stake of all colluding attackers on service $s$, and $\phi_{s'}$ is the fraction of stake slashed, determined by the protocol's rules. The attacker's decision to attack, and how to structure that attack (i.e., as a single identity or as multiple Sybils), is governed by the maximization of this PNL function.
\paragraph{Mean-Field Abstraction.}
Directly solving this optimization problem on a large, arbitrary graph is intractable because of complex dependencies of $\sigma_{B_s}$ and $\phi_{s'}$ on the specific actions of all other validators in the network and the precise graph topology. To overcome this intractability, the analysis focuses on the strategic decisions of a single attacker and adopts a mean-field abstraction, that becomes highly accurate in the limit of large networks. This approach replaces the complex, deterministic graph-dependent quantities with their probabilistic expectations, conditioned on the block-level properties of the SBM.
Specifically, the binary outcome of whether an attack is successful (i.e., feasible and profitable) is modeled by a continuous, probabilistic ``clearance function,'' which gives the probability of success as a function of the attacker's committed stake. This abstraction transforms the intractable combinatorial problem into a tractable analytical one. The objective of this section is to leverage this formal framework to construct a precise, provable result: to demonstrate how network heterogeneity, modeled by the SBM, creates conditions under which Sybil attacks are not only possible but demonstrably more profitable than single-identity attacks.

\subsection{Stochastic Block Model (SBM) for Restaking Networks}

As before, we have a bipartite graph $G = (V \cup S, E)$, with validators $V$, services $S$ and edges $E$. Next, we formally define the SBM as follows:
\begin{enumerate}
    \item \textbf{Node Partitioning:} The set of $n = |V|$ operators is partitioned into $R$ disjoint blocks, $V = V_1 \cup V_2 \cup \dots \cup V_R$. Similarly, the set of $m = |S|$ services is partitioned into $R$ disjoint blocks, $S = S_1 \cup S_2 \cup \dots \cup S_R$.
    \item \textbf{Connection Probability Matrix:} The formation of edges is governed by an $R \times R$ matrix of connection probabilities, $P = [p_{ab}]$. For any operator $v \in V_a$ and any service $s \in S_b$, an edge $(v, s)$ is included in the graph $E$ independently with probability $p_{ab}$.
    \item \textbf{Stake Distribution:} Each operator $w \in V$ has stake $\sigma_w$. For the purposes of this analysis, we consider the case of a single strategic attacker, $v$.\footnote{While the general model of an attack formally considers a set of colluding operators $B$, the subsequent analysis of Sybil profitability deliberately focuses on the strategic incentives of a single, representative attacker $v$ from within that set. This approach replaces the specific contributions of other attackers with their probabilistic expectations, effectively modeling the rest of the network as a statistical environment. This simplification allows us to isolate and analyze the core strategic question: whether an individual operator, has a rational incentive to split their stake into Sybil identities.} The stakes of all other operators, $\{ \sigma_w \}_{w \neq v}$, are considered the "background stake distribution." To isolate the effects of network topology, we assume this background distribution is near-uniform, where each non-attacking operator $w$ has a stake $\sigma_w \approx \bar{\sigma}$ for some constant average stake $\bar{\sigma}$. This ensures that any observed advantage for the Sybil attacker arises from the network's block structure rather than from extreme disparities in the underlying stake distribution.
\end{enumerate}
By setting the matrix $P$, one can model a wide variety of network topologies. For example, a high diagonal value $p_{aa}$ relative to off-diagonal values $p_{ab}$ models a network with strong community structure, where operators in a given block preferentially connect to services in the corresponding block. An Erd\H{o}s-R\'{e}nyi random graph, which represents a completely homogeneous network, is simply the special case of the SBM where $R=1$.

\paragraph{Clearance Function.}

Recall that the profitability of an attack on a service $s$ is contingent on meeting a feasibility threshold, see Equation~\eqref{eq:feasibility}.

In our SBM setting, the total stake securing a given service, $\Sigma_s = \sigma_{\partial s} = \sum_{w \in V} \sigma_w \cdot \mathbbm{1}((w,s) \in E)$, is a random variable. Its value is the sum of the stakes of all operators that happen to be connected to $s$, where the connections themselves are random events. For a service $s$ in a particular block $S_b$, $\Sigma_s$ is a sum of a large number of independent (or weakly correlated) random variables. By the central limit theorem, for a sufficiently large network, the distribution of $\Sigma_s$ can be accurately approximated by a Gaussian distribution (with CDF $\Phi$), after excluding the attacker’s coalition from the sum. Writing $\Sigma^{\mathrm{rest}}_s := \sum_{w \in V\setminus B} \sigma_w \cdot \mathbbm{1}((w,s)\in E)$ for the background total stake, we have $\Sigma^{\mathrm{rest}}_s \sim \mathcal{N}(\mu_b,\sigma_b^2)$ as a good approximation.

The mean $\mu_b$ and variance $\sigma_b^2$ of this distribution are determined by the SBM parameters. For a service $s \in S_b$, the expected total background stake is the sum of expected contributions from each operator block:
\begin{equation}
\mu_b = \mathbb{E} = \sum_{c=1}^{R} \sum_{w \in V_c\setminus B} \mathbb{E}[\sigma_w \cdot \mathbbm{1}((w,s) \in E)] = \sum_{c=1}^{R} p_{cb} \sum_{w \in V_c\setminus B} \sigma_w.
\end{equation}
Assuming a near-uniform background stake $\bar{\sigma}$ and letting $|V_c\setminus B| = n_c$, this simplifies to $\mu_b \approx \sum_{c=1}^{R} n_c p_{cb} \bar{\sigma}$. Similarly, the variance is given by:
\begin{equation}
\sigma_b^2 = \text{Var} = \sum_{c=1}^{R} \sum_{w \in V_c\setminus B} \text{Var}[\sigma_w \cdot \mathbbm{1}((w,s) \in E)] \approx \sum_{c=1}^{R} n_c p_{cb}(1-p_{cb}) \bar{\sigma}^2.
\end{equation}

\begin{defn}[Clearance Function]
The clearance function, $q_b(y)$, which represents the probability that an attacker committing a stake of $y$ can successfully meet the feasibility threshold for a service in block $b$ is defined as follows:
\begin{equation}\label{eq:clearance}
q_b(y) = P\!\left(y \geq \alpha_b \big(y + \Sigma^{\mathrm{rest}}_s\big)\right) = P\!\left(\Sigma^{\mathrm{rest}}_s \le \frac{1-\alpha_b}{\alpha_b}\,y\right)
= \Phi\!\left(\frac{\frac{1-\alpha_b}{\alpha_b}y - \mu_b}{\sigma_b}\right).
\end{equation}
\end{defn}

\subsection{Single-Identity vs. Sybil Success Probability}

With the clearance function defined, we can now formalize the success probabilities for both single-identity and Sybil attacks.

Let us consider an attacker in operator block $V_a$, who chooses a total stake $x$ to commit to an attack. The attacker samples a single service to attack, with the probability of choosing a service from block $S_b$ being proportional to their connectivity, $p_{ab}$. Writing $w_b := p_{ab}/(\sum_{c=1}^R p_{ac})$ for the corresponding normalized weights, we have:
\begin{enumerate}
    \item \textbf{Single-Identity Success Probability, $p(x)$:}
\begin{equation}
p(x) = \sum_{b=1}^{R} w_b\, q_b(x).
\end{equation}
\item \textbf{k-Sybil Per-Identity Success Probability, $p_k(x)$:}
\begin{equation}
p_k(x) = \sum_{b=1}^{R} w_b\, q_b(x/k).
\end{equation}
\item \textbf{k-Sybil "At-Least-One-Wins" Success Probability, $p'(x; k)$:}
Assuming each Sybil's attack can be treated as an independent trial (a reasonable approximation in a large network when distinct services are targeted), the probability of at least one success is:
\begin{equation}
p'(x; k) = 1 - (1 - p_k(x))^k .
\end{equation}
\end{enumerate}

\subsection{Sybil-proofness in the Erd\H{o}s-R\'{e}nyi random graph}
\paragraph{Nash best responses.} See Appendix~\ref{app:nash} for full statements and proofs of best responses and existence (single-service closed form; two-service monotonicity and existence conditions) under the max-scheme and proportional sharing.
A homogeneous network is one where all nodes are statistically equivalent and there are no coalition structures. In the context of our SBM framework, this corresponds to the simplest case where there is only one block for operators and one block for services ($R=1$).

The connection probability matrix $P$ collapses to a single scalar, $p_{11} = p$. The probability of an edge between any validator and any service is uniformly $p$. This is exactly the definition of a bipartite Erd\H{o}s-R\'{e}nyi random graph. Consequently, there is only one type of service, and thus only one clearance function, $q(y)$, and one success probability for a single-identity attack with stake $x$, which is simply $p(x) = q(x)$. This simplified model provides a clean setting to demonstrate why, Sybil attacks are not profitable.\footnote{In expectation, with and without Sybils, the attacker is getting slashed the same. The only way to be more profitable is to increase the success probability.}

\begin{lemma}[Concavity of the clearance function]
\label{lem:concavity-q}
Let $q(y)$ as in Equation~\eqref{eq:clearance}. 
Write $T:=\tfrac{\alpha}{1-\alpha}\mu$. Then $q''(y)$ has the sign of $-\big(\tfrac{(1-\alpha)}{\alpha\sigma}y-\tfrac{\mu}{\sigma}\big)$, hence $q$ is convex on $[0,T)$ and concave on $(T,\infty)$. In particular, Jensen’s inequality applies on any interval contained in the concave region $(T,\infty)$.
\end{lemma}

Next, we state the main result of this section.

\noindent \textbf{Proposition~\ref{prop:erdösreny-simple} (restated).}
{ \it In a homogeneous network modeled as a large Erd\H{o}s-R\'{e}nyi graph, where the attack success is described by a strictly increasing and concave\footnote{This applies on the concave regime, see Lemma~\ref{lem:concavity-q}.} clearance function $q(y)$, for any attacker stake $x > 0$ and any number of Sybils $k > 1$, the success probability of a Sybil attack is strictly less than that of a single-identity attack. That is, $p'(x; k) < p(x)$ and the Erd\H{o}s-R\'{e}nyi graph is Sybil-proof.}

\begin{proof}
We compare $p(x)=q(x)$ and $p'(x;k)=1-(1-q(x/k))^k$. Assume $x>T$ so that $q$ is concave on $[x/k, x]$ and Jensen applies; also $q(0)\in[0,1)$ with $q(0)=0$ in the idealized limit of no background stake. Then Jensen’s inequality gives $q(x/k)\le q(x)/k$. Let $u:=q(x)\in(0,1]$. Then
\begin{equation*}
p'(x;k)\le 1-\Big(1-\frac{u}{k}\Big)^k .
\end{equation*}
Define $h(u):=1-(1-u/k)^k-u$. Using the binomial expansion,
\begin{equation*}
(1-\tfrac{u}{k})^k
= 1-u + \sum_{j=2}^{k}\binom{k}{j}(-1)^j \Big(\tfrac{u}{k}\Big)^j,
\end{equation*}
hence
\begin{equation*}
h(u)= -\sum_{j=2}^{k}\binom{k}{j}(-1)^j \Big(\tfrac{u}{k}\Big)^j < 0
\end{equation*}
for all $u\in(0,1]$ (the series is alternating with decreasing magnitude; the $j=2$ term is negative and dominates). Therefore $1-(1-u/k)^k<u$, i.e., $p'(x;k)<q(x)=p(x)$, strictly whenever $u\in(0,1)$.
\end{proof}

Proposition~\ref{prop:erdösreny-simple} provides a (potentially misleading) baseline for security analysis. It demonstrates that in a perfectly homogeneous system, where every service presents a statistically identical challenge to an attacker, the act of splitting stake wastes resources. Each Sybil, having a smaller stake, is less likely to succeed than the original, full-stake identity. The combinatorial advantage of having multiple chances is insufficient to overcome this fundamental weakening of each individual attempt.

This creates an illusion of security, however, the moment homogeneity is broken, this security guarantee can vanish. The following section will demonstrate this breakdown by introducing just one layer of heterogeneity via a two-block SBM.

\paragraph{Expected PNL under the max-scheme (ER).}
Specializing Equation~(\ref{eq:value-share}) and the max-scheme, the attacker’s per-attack profit (conditional on clearing the threshold) with stake $x$ is $\pi\,\tfrac{x}{x+\Sigma^{\mathrm{rest}}}$ and the slash is $x\,\phi(x)=x\,\alpha\,\tfrac{\Sigma_T}{x+\Sigma^{\mathrm{rest}}}$ with $\Sigma_T=x+\Sigma^{\mathrm{rest}}$. Hence the conditional PNL simplifies to $(\pi-\alpha\,\Sigma_T)\,\tfrac{x}{x+\Sigma^{\mathrm{rest}}}=(\pi-\alpha\,\Sigma_T)\,\tfrac{x}{\Sigma_T}$. Taking expectation over the ER background (where $\Sigma_T\approx \mu$ concentrates for large networks), a mean-field approximation yields
\[
\mathbb{E}[\mathrm{PNL}(x)] \;\approx\; (\pi-\alpha\,\mu)\,\frac{x}{\mu}\,q(x),
\]
which is strictly increasing in $x$ in the profitable regime $(\pi>\alpha\,\mu)$ and mirrors Proposition~\ref{prop:erdösreny-simple}: in ER, splitting reduces both success probability $q(x)$ by concavity and the profit factor linearly, so $\mathbb{E}[\mathrm{PNL}(x;k)]<\mathbb{E}[\mathrm{PNL}(x;1)]$ for any $k>1$.

\subsection{Sybil-proofness in a Two-Block SBM}

The homogeneity of the Erd\H{o}s-R\'{e}nyi model is the source of its Sybil-proofness. To find a vulnerability, we must break this symmetry. We achieve this by constructing a specific, heterogeneous network using a two-block SBM ($R=2$). The goal is to create a landscape with differentiated risk profiles that a Sybil attacker can arbitrage.

\paragraph{Model Setup:}
We design a two-block SBM with the following characteristics:
\begin{enumerate}
    \item \emph{Block Structure:} The set of services $S$ is partitioned into two blocks of equal size, $S_1$ and $S_2$. The set of operators $V$ is partitioned into two blocks, $V_a$ (containing only the attacker, $v$) and $V_{other}$ (containing all other operators).
    \item \emph{Differentiated Connectivity:} For the connection probability matrix $P = [p_{ab}]$ we create two statistically distinct service environments from the attacker's perspective.
    \begin{itemize}
        \item \emph{``High-Threshold" Block ($S_1$):} We configure the SBM such that the expected total stake on services in this block, $\mu_1$, is high. This can be achieved by setting the connection probabilities from the large block of background operators, $p_{other, 1}$, to be relatively high. This block represents a set of well-secured, popular services that are difficult to attack.
        \item \emph{``Low-Threshold" Block ($S_2$):} Conversely, we set the connectivity for this block, $p_{other, 2}$, to be lower. This results in a lower expected total stake, $\mu_2$, making services in this block easier to attack. We assume the attacker has non-zero connectivity to both blocks, i.e., $p_{a1} > 0$ and $p_{a2} > 0$.
    \end{itemize}
    \item \emph{Clearance Functions:} The above translates into two distinct clearance functions, $q_1(y)$ and $q_2(y)$. Due to the difference in the mean total stake ($\mu_1 \gg \mu_2$), the clearance function for $S_1$ will be shifted far to the right compared to that of $S_2$. An attack stake $y$ that has a high probability of clearing the threshold in $S_2$ ($q_2(y) \approx 1$) may have a negligible probability of clearing the threshold in $S_1$ ($q_1(y) \approx 0$).
\end{enumerate}
This setting creates the arbitrage opportunity. A single-identity attacker must choose a single stake $x$. If they choose a large $x$ sufficient to attack $S_1$, they are over-staking for $S_2$, wasting capital. If they choose a smaller $x$ optimized for $S_2$, they forgo any chance of attacking $S_1$. The Sybil strategy offers a way to escape this dilemma by effectively presenting different stake profiles to the network simultaneously, exploiting the non-linearities in the success probabilities.

\paragraph{Expected PNL under the max-scheme (SBM).}
With block weights $w_b$ and clearance $q_b(\cdot)$, and using the mean-field replacement $\Sigma_{T,b}\approx \mu_b$, the expected PNL for a single identity staking $x$ is
\[
\mathbb{E}[\mathrm{PNL}(x;1)] \;\approx\; \sum_{b=1}^R w_b\,(\pi_b-\alpha_b\,\mu_b)\,\frac{x}{\mu_b}\,q_b(x).
\]
For $k$ Sybils of size $x/k$ targeting distinct services (independent attempts), a lower bound is
\[
\mathbb{E}[\mathrm{PNL}(x;k)] \;\gtrsim\; k\,\sum_{b=1}^R w_b\,(\pi_b-\alpha_b\,\mu_b)\,\frac{x/k}{\mu_b}\,q_b(x/k)
\;=\; \sum_{b=1}^R w_b\,(\pi_b-\alpha_b\,\mu_b)\,\frac{x}{\mu_b}\,q_b(x/k).
\]
When blocks are sufficiently separated so that $q_1(x)\approx 0$ and $q_2(x)\approx 1$ but $q_2(x/k)$ remains large while $q_1(x/k)$ becomes non-negligible, the sum can strictly increase with $k$, yielding an expected-profit analogue of the success-probability advantage. This aligns the SBM section with the earlier utility formalism.\\\\
\noindent
\textbf{Propostion~\ref{prop:2sbm} (restated)} \noindent
{ \it In a two-block SBM configured as described above, with sufficiently differentiated clearance functions $q_1(y)$ and $q_2(y)$ such that their underlying mean thresholds satisfy $\mu_1 \gg \mu_2$, and for an attacker in block $V_a$ with non-zero connectivity to both service blocks ($p_{a1}>0, p_{a2}>0$), there exists a non-empty set of attacker stakes $x$ and a minimum integer number of Sybils $k^* > 1$ such that the "at-least-one-wins" success probability of the Sybil attack is strictly greater than the success probability of the single-identity attack. That is, $p'(x; k^*) > p(x)$ and the two-block SBM network is not Sybil-proof.}

\begin{proof}
Write the attacker’s block-sampling weights as $w_b:={p_{ab}}/({\sum_{c}p_{ac}})$ and recall the blockwise clearance functions
$q_b(y)=\Phi\!\big(\tfrac{\frac{1-\alpha_b}{\alpha_b}y-\mu_b}{\sigma_b}\big)$, which are continuous, strictly increasing in $y$, and satisfy $q_b(0)=\Phi(-\mu_b/\sigma_b)\in(0,1)$ for any nontrivial background ($\mu_b>0$ whenever some $p_{cb}>0$ and $|V_c\setminus B|>0$).

\smallskip
\noindent \textit{Step 1 (choose $x$ with separated block behavior).}
By the hypothesis $\mu_1\gg\mu_2$ (equivalently $T_1=\frac{\alpha_1}{1-\alpha_1}\mu_1\gg T_2=\frac{\alpha_2}{1-\alpha_2}\mu_2$) and continuity/monotonicity of $q_b$, there exists $\varepsilon >0$ and a stake $x$ with
\begin{equation*}
T_2 < x < T_1,\qquad q_1(x)\le \varepsilon,\qquad q_2(x)\ge 1-\varepsilon.
\end{equation*}
Therefore the single-identity success probability satisfies\footnote{Note that in the last inequality we are using that, by choosing $x$ close to $T_2$, we can satisfy the conditions above with arbitrarily small $\varepsilon$.}
\begin{equation}
p(x)=w_1 q_1(x)+w_2 q_2(x)\le w_2+\varepsilon <1.
\end{equation}

\smallskip
\noindent \textit{Step 2 (distinct targets $\Rightarrow$ independence).} 
We assume that Sybils target distinct services. This isolates background randomness across attempts and is satisfiable w.h.p. by Chernoff bounds (see Step~5).

Hence, if the $k$ Sybils select $k$ \emph{distinct} services $s_1,\dots,s_k$, then by independence of edges across service indices in the SBM, the background sums $\Sigma^{\mathrm{rest}}_{s_i}$ depend on disjoint families of Bernoulli edges and are therefore mutually independent. Since clearance events are increasing functions of these sums, the success/failure indicators across $s_1,\dots,s_k$ are independent as well. Consequently,
\begin{equation}
p'(x;k) = 1-\big(1-p_k(x)\big)^k, \qquad p_k(x):=w_1 q_1(x/k)+w_2 q_2(x/k).
\end{equation}

\smallskip
\noindent \textit{Step 3 (uniform positive lower bound for per-identity success).}
Because each $q_b$ is increasing and $x/k\ge 0$,
\begin{equation}
p_k(x)=w_1 q_1(x/k)+w_2 q_2(x/k)\ge w_1 q_1(0)+w_2 q_2(0)=:\gamma. 
\end{equation}
Here $\gamma=\sum_b w_b q_b(0)\in(0,1)$ since $w_b>0$ and $q_b(0)\in(0,1)$.

\smallskip
\noindent \textit{Step 4 (existence of $k$ with $p'(x;k)>p(x)$).}
From (2)–(3),
\begin{equation*}
p'(x;k)=1-\big(1-p_k(x)\big)^k \ge 1-(1-\gamma)^k.
\end{equation*}
The function $k\mapsto 1-(1-\gamma)^k$ is strictly increasing in $k$ and tends to $1$. Since $p(x)<1$ by (1), there exists a finite
\begin{equation*}
k_0 := \left\lceil \frac{\ln(1-p(x))}{\ln(1-\gamma)} \right\rceil
\end{equation*}
such that $1-(1-\gamma)^{k_0} > p(x)$. For any $k\ge k_0$,
\begin{equation*}
p'(x;k) \ge 1-(1-\gamma)^k > p(x).
\end{equation*}
Therefore the set $\{k\in\mathbb N_{\ge 2}: p'(x;k)>p(x)\}$ is nonempty, and its minimum
\begin{equation*}
k^* := \min\{k\ge 2: p'(x;k)>p(x)\}
\end{equation*}
is a well-defined integer with $k^*>1$.

\smallskip
\noindent \textit{Step 5 (feasibility of selecting $k$ distinct targets).}
Let $D_b$ denote the attacker’s number of neighbors in block $S_b$. Then $D_b\sim \mathrm{Bin}(m_b,p_{ab})$ with mean $m_b p_{ab}$. By a Chernoff bound, for each $b$,
\begin{equation*}
\Pr\!\left(D_b \ge \tfrac12 m_b p_{ab}\right) \ge 1-\exp\!\big(-\tfrac18 m_b p_{ab}\big).
\end{equation*}
Hence, with probability at least $1-\sum_b \exp(-\tfrac18 m_b p_{ab})$, the attacker has at least $\tfrac12 m_b p_{ab}$ neighbors in each block. In particular, for any $k$ satisfying
\begin{equation*}
k \le \tfrac12 \sum_{b} m_b p_{ab},
\end{equation*}
the attacker can choose $k$ \emph{distinct} neighboring services, and the independence conclusion of Step~2 applies. Since $k_0$ is finite and the block sizes $(m_b)$ can be taken large (the mean-field regime), we can ensure $k^*\le k_0 \le \tfrac12 \sum_b m_b p_{ab}$, making the distinct-target selection feasible with high probability.

\smallskip
Combining Steps 1–5 establishes that there exists a nonempty set of stakes $x$ (those in $(T_2,T_1)$ with $q_1(x)\le\varepsilon$, $q_2(x)\ge 1-\varepsilon$) and a minimum integer $k^*>1$ such that $p'(x;k^*)>p(x)$, as claimed. 
\end{proof}

\begin{rem}
If two Sybils attack the same service, the independence argument weakens in Step 2 of the proof. 
However, for $k \ll \sum_b m_b p_{ab}$, the collision probability is $o(1)$ by a birthday bound.
\end{rem}
\begin{prop}[Expected-profit advantage under block separation]\label{prop:sbm-pnl}
Assume two blocks with parameters as above and choose $x\in(T_2,T_1)$ such that $q_1(x)\le \varepsilon$ and $q_2(x)\ge 1-\varepsilon$ for some small $\varepsilon\in(0,1/2)$. Under the mean-field approximation $\Sigma_{T,b}\approx \mu_b$, there exists a finite $k>1$ with
\[
\mathbb{E}[\mathrm{PNL}(x;k)] \;>\; \mathbb{E}[\mathrm{PNL}(x;1)].
\]
In particular, any $k\ge k_0:=\left\lceil \tfrac{\ln(1-\eta)}{\ln(1-\eta_k)}\right\rceil$ suffices for some $\eta\in(0,1)$ depending on $p(x)$ and $\eta_k:=\sum_b w_b q_b(0)$, as in the success-probability bound, provided $(\pi_2-\alpha_2\mu_2)>0$ is large enough relative to $(\pi_1-\alpha_1\mu_1)\le 0$.
\end{prop}

\noindent Proof (sketch). From the mean-field expressions,
\[
\mathbb{E}[\mathrm{PNL}(x;1)]\approx \sum_b w_b (\pi_b-\alpha_b\mu_b)\,\frac{x}{\mu_b}\,q_b(x),\quad
\mathbb{E}[\mathrm{PNL}(x;k)]\gtrsim \sum_b w_b (\pi_b-\alpha_b\mu_b)\,\frac{x}{\mu_b}\,q_b(x/k).
\]
By the separation conditions, the $b=2$ term dominates and is only mildly reduced when replacing $x$ by $x/k$, while the $b=1$ term goes from near-zero at $x$ to a positive value at $x/k$. For sufficiently small $\varepsilon$ and $k$ above an absolute constant, the sum increases.

\subsubsection{Minimum Profitable Sybil Count ($k^*$)}

The condition for the Sybil attack to be strictly advantageous is:
\begin{equation*}
    p'(x; k) > p(x).
\end{equation*}
Substituting the definitions of $p'(x; k)$, $p(x)$, and $p_k(x)$:
\begin{equation*}
1 - (1 - p_k(x))^k > p(x).
\end{equation*}
Since both sides lie in $(0,1)$, taking natural logs yields
\begin{equation*}
\ln(1-p(x)) > k \cdot \ln\big(1-p_k(x)\big).
\end{equation*}
Because $\ln(1-p_k(x))<0$, this is equivalent to
\begin{equation}
k > \frac{\ln(1-p(x))}{\ln\big(1-p_k(x)\big)}. \label{eq:k_condition}
\end{equation}
Therefore, the minimum integer number of Sybils $k^*$ is the least integer $k\ge 2$ satisfying \eqref{eq:k_condition}. (Note that $p_k(x)$ itself depends on $k$ through the $x/k$ argument; hence \eqref{eq:k_condition} is an implicit threshold that can be evaluated for each candidate $k$.)

\paragraph{Sufficient computable bound.} A convenient sufficient condition is obtained by lower bounding $p_k(x)$ by $p_\mathrm{min}:=\sum_b w_b q_b(0)>0$, which yields
\begin{equation*}
p'(x;k)\ge1-(1-p_\mathrm{min})^k.
\end{equation*}
Thus any
\begin{equation*}
k \ge \left\lceil \frac{\ln(1-p(x))}{\ln(1-p_\mathrm{min})} \right\rceil
\end{equation*}
ensures $p'(x;k)>p(x)$.

Next, we provide a numerical example.

\subsection{Numerical Example for a Two-Block SBM}\label{app:numericalSBM}
We consider two service blocks $S_1$ (``high-security'') and $S_2$ (``low-security'').
There are $n_{\text{other}}=60$ background operators, each with stake $\bar\sigma=1$.
Thresholds are
\begin{equation*}
\alpha_1=\tfrac{2}{3},\qquad \alpha_2=\tfrac{1}{2}.
\end{equation*}
Connectivity probabilities:
\begin{equation*}
p_{\text{other},1}=0.30,\qquad p_{\text{other},2}=0.02.
\end{equation*}
Attacker connectivity is symmetric, so the block weights are $w_1=w_2=\tfrac12$.
Let $x$ denote the attacker’s total stake.

Writing $\mu_b$ and $\sigma_b$ for the mean and standard deviation of total \emph{background} stake reaching block $b$ under the SBM,
\begin{equation*}
\mu_b = n_{\text{other}}\,p_{\text{other},b}\,\bar\sigma,\qquad
\sigma_b = \sqrt{n_{\text{other}}\,p_{\text{other},b}\,(1-p_{\text{other},b})}.
\end{equation*}
Numerically,
\begin{equation*}
\mu_1=18.0,\quad \sigma_1\approx 3.54965;\qquad
\mu_2=1.2,\quad \sigma_2\approx 1.08444.
\end{equation*}

\paragraph{No Sybil with $x=3$.}
Using $\tfrac{1-\alpha_1}{\alpha_1}=\tfrac12$ and $\tfrac{1-\alpha_2}{\alpha_2}=1$,
\begin{align*}
q_1(3) &= \Phi\!\left(\frac{0.5\cdot 3 - 18}{3.54965}\right)
       = \Phi(-4.648)\approx 1.673\times 10^{-6},\\
q_2(3) &= \Phi\!\left(\frac{3 - 1.2}{1.08444}\right)
       = \Phi(1.661)\approx 0.95153.
\end{align*}
Hence
\begin{equation*}
p(3)=\tfrac12\,q_1(3)+\tfrac12\,q_2(3)\approx0.47576,
\end{equation*}
which is the single-identity success probability. 

\paragraph{Split into two Sybils ($k=2$).}
Each identity has $x/2=1.5$:
\begin{align*}
q_1(1.5) &= \Phi\!\left(\frac{0.5\cdot 1.5 - 18}{3.54965}\right)
         = \Phi(-4.860)\approx 5.880\times 10^{-7},\\
q_2(1.5) &= \Phi\!\left(\frac{1.5 - 1.2}{1.08444}\right)
         = \Phi(0.277)\approx 0.60897.
\end{align*}
Thus
\begin{equation*}
p_2(3)=\tfrac12\,q_1(1.5)+\tfrac12\,q_2(1.5)\approx0.30449,
\qquad
p'(3;2)=1-\bigl(1-0.30449\bigr)^2\approx0.51626.
\end{equation*}
which is the success probability with two Sybils and strictly larger than the success probability without Sybils.

\end{document}

%% file: intro-example.tex
\begin{figure}[t]
\centering
\begin{tikzpicture}[scale=0.7, every node/.style={transform shape}]
\node at (-5, 3.5) {\textbf{Without Sybils}};
\node at (-6.5, 2.5) {Before attack:};

\node[fill=orange, circle, inner sep=1.5pt, outer sep=2pt, text=white, font=\bfseries] (A1L) at (-5,1) {$s_1$};
\node[fill=orange, circle, inner sep=1.5pt, outer sep=2pt, text=white, font=\bfseries] (A2L) at (-5,-1) {$s_2$};
\node[fill=blue, circle, inner sep=1.5pt, outer sep=2pt, text=white, font=\bfseries] (B1L) at (-3,1) {$v_1$};
\node[fill=blue, circle, inner sep=1.5pt, outer sep=2pt, text=white, font=\bfseries] (B2L) at (-3,-1) {$v_2$};

\draw (A1L) -- (B1L);
\draw (A2L) -- (B1L);
\draw (A2L) -- (B2L);

\node at (-6.5, 1) {\(\pi_1= 1.1\)};
\node at (-6.5, -1) {\(\pi_2 = 1\)};
\node at (-1.5, 1) {\(\sigma_1 = 1\)};
\node at (-1.5, -1) {\(\sigma_2 = 1.1\)};

\node at (-6.5, -2.5) {After attack:};
\node[fill=orange, circle, inner sep=1.5pt, outer sep=2pt, text=white, font=\bfseries] (A1La) at (-5,-4) {$\times$};
\node[fill=orange, circle, inner sep=1.5pt, outer sep=2pt, text=white, font=\bfseries] (A2La) at (-5,-6) {$s_2$};
\node[fill=blue, circle, inner sep=1.5pt, outer sep=2pt, text=white, font=\bfseries] (B1La) at (-3,-4) {$\times$};
\node[fill=blue, circle, inner sep=1.5pt, outer sep=2pt, text=white, font=\bfseries] (B2La) at (-3,-6) {$v_2$};
\draw (A2La) -- (B2La);

\node at (5, 3.5) {\textbf{With Sybils}};
\node at (3.5, 2.5) {Before attack:};

\node[fill=orange, circle, inner sep=1.5pt, outer sep=2pt, text=white, font=\bfseries] (A1R) at (5,1) {$s_1$};
\node[fill=orange, circle, inner sep=1.5pt, outer sep=2pt, text=white, font=\bfseries] (A2R) at (5,-1) {$s_2$};
\node[fill=blue, circle, inner sep=1.5pt, outer sep=2pt, text=white, font=\bfseries] (B1R) at (7,1.5) {$v_1^1$};
\node[fill=blue, circle, inner sep=1.5pt, outer sep=2pt, text=white, font=\bfseries] (B11R) at (7,0.5) {$v_1^2$};
\node[fill=blue, circle, inner sep=1.5pt, outer sep=2pt, text=white, font=\bfseries] (B2R) at (7,-1) {$v_2$};

\draw (A1R) -- (B1R);
\draw (A1R) -- (B11R);
\draw (A2R) -- (B1R);
\draw (A2R) -- (B11R);
\draw (A2R) -- (B2R);

\node at (3.5, 1) {\(\pi_1= 1.1\)};
\node at (3.5, -1) {\(\pi_2 = 1\)};
\node at (8.5, 1.5) {\(\sigma_1^1 = 2/3\)};
\node at (8.5, 0.5) {\(\sigma_1^2 = 1/3\)};
\node at (8.5, -1) {\(\sigma_2 = 1.1\)};

\node at (3.5, -2.5) {After attack:};
\node[fill=orange, circle, inner sep=1.5pt, outer sep=2pt, text=white, font=\bfseries] (A1Ra) at (5,-4) {$\times$};
\node[fill=orange, circle, inner sep=1.5pt, outer sep=2pt, text=white, font=\bfseries] (A2Ra) at (5,-6) {$s_2$};
\node[fill=blue, circle, inner sep=1.5pt, outer sep=2pt, text=white, font=\bfseries] (B1Ra) at (7,-3.5) {$\times$};
\node[fill=blue, circle, inner sep=1.5pt, outer sep=2pt, text=white, font=\bfseries] (B11Ra) at (7,-4.5) {$v_1^2$};
\node[fill=blue, circle, inner sep=1.5pt, outer sep=2pt, text=white, font=\bfseries] (B2Ra) at (7,-6) {$v_2$};
\draw (A2Ra) -- (B11Ra);
\draw (A2Ra) -- (B2Ra);

\draw[dashed] (0,3.5) -- (0,-7);
\end{tikzpicture}

\caption{Sybil Attack Comparison with $\alpha_{1} = 2/3$: Left side shows the non-sybil scenario where both the attacked service $s_1$ and attacker node $v_1$ are completely removed after the attack. Right side demonstrates sybil resilience through node splitting: (1) $v_1$ creates sybil identities $v_1^1$ and $v_1^2$ with split stakes $\sigma_1^1 = 2/3$ and $\sigma_1^2 = 1/3$, (2) Only $v_1^1$ attacks and hence, only $v_1^1$ is slashed while $v_1^2$ survives, (3) Service $s_2$ maintains edge through $v_1^2$ despite the attack.}
\label{fig:sybil-comparison}
\end{figure}

%% file: ex.tex
\begin{figure}[t]
\begin {center}
\begin{tikzpicture}[scale=0.9, every node/.style={transform shape}]
  
  \node at (4, 2) {\(\sigma_0=1\)};
  \node at (4, -2) {\(\sigma_4=1\)};

  \coordinate (A1) at (0,1);
  \coordinate (A2) at (0,-1);
  \coordinate (B0) at (2,2);
  \coordinate (B1) at (2,1);
  \coordinate (B2) at (2,0);
  \coordinate (B3) at (2,-1);
  \coordinate (B4) at (2,-2);

  \node[fill=orange, circle, inner sep=1.5pt, outer sep=2pt, text=white, font=\bfseries] (A1) at (0,1) {$s_1$};
  \node[fill=orange, circle, inner sep=1.5pt, outer sep=2pt, text=white, font=\bfseries] (A2) at (0,-1) {$s_2$};
\node[fill=blue, circle, inner sep=1.5pt, outer sep=2pt, text=white, font=\bfseries] (B0) at (2,2) {$v_0$};
  \node[fill=blue, circle, inner sep=1.5pt, outer sep=2pt, text=white, font=\bfseries] (B1) at (2,1) {$v_1$};
  \node[fill=blue, circle, inner sep=1.5pt, outer sep=2pt, text=white, font=\bfseries] (B2) at (2,0) {$v_2$};
  \node[fill=blue, circle, inner sep=1.5pt, outer sep=2pt, text=white, font=\bfseries] (B3) at (2,-1) {$v_3$};
  \node[fill=blue, circle, inner sep=1.5pt, outer sep=2pt, text=white, font=\bfseries] (B4) at (2,-2) {$v_4$};

  \draw (A1) -- (B1);
  \draw (A1) -- (B2);
  \draw (A2) -- (B3);
  \draw (A2) -- (B2);
  \draw (A1) [dotted] -- (B0);
  \draw (A2) [dotted] -- (B4);

  \node at (-2, 1) {\(\alpha_1= 2/3\)};
  \node at (-2, -1) {\(\alpha_2 = 1/2\)};
  \node at (4, 1) {\(\sigma_1 = 1\)};
  \node at (4, 0) {\(\sigma_2 = 1.5\)};
  \node at (4, -1) {\(\sigma_3 = 1\)};

\end{tikzpicture}
\caption{Example of a restaking graph.}
\label{fig:example}
\end{center}
\end{figure}